\documentclass[preprint2]{aastex}

\newcommand{\EWmin}{EW$_{min}$}
\begin{document}
\title{ESO-VLT optical spectroscopy of BL Lac objects: II. New redshifts, 
featureless objects and classification assessments.}
\author{B. Sbarufatti\altaffilmark{1}, A. Treves}
\affil{Universit\`a dell'Insubria, Via Valleggio 11, I-22100 Como, Italy}
\author{R. Falomo}
\affil{INAF, Osservatorio Astronomico di Padova, Vicolo dell'Osservatorio 5, 
 I-35122 Padova, Italy}
\author{J. Heidt}
\affil{Landenssterwarte Heidelberg, K\"onigstuhl, D-69117 Heidelberg, Germany}
\author{J. Kotilainen}
\affil{Tuorla Observatory, University of Turku, V\"ais\"al\"antie 20, FIN-21500
 Piikki\"o, Finland}
\author{R. Scarpa}
\affil{European Southern Observatory, 3107 Alonso de Cordova, Santiago, Chile}
\altaffiltext{1}{also at Universit\`a di Milano-Bicocca}
\shorttitle{ESO-VLT optical spectroscopy of BL Lac objects: II}
\shortauthors{Sbarufatti et al.}
\begin{abstract}
We report on ESO Very Large Telescope optical spectroscopy 
of 42 BL Lacertae objects of unknown redshift.  Nuclear emission lines 
were observed in 12 objects, while for another six we detected 
absorption features due to their host galaxy. The new high S/N spectra 
therefore allow us to measure the redshift of 18 sources. Five of the
observed objects were reclassified either as stars or quasars, and 
one is of uncertain nature.  For the remaining 18 the optical 
spectra  appear without intrinsic features in spite of our 
ability to measure rather faint (EW $\sim$0.1 \AA) spectral lines.
For the latter sources a lower limit to the redshift was set exploiting the 
very fact that the absorption lines of the host galaxy are 
undetected on the observed spectra.

\end{abstract}


\section{Introduction}\label{sec:intro}

BL Lac objects (hereinafter BL Lacs or BLL) are active galactic nuclei 
(AGN) characterized by luminous, rapidly variable UV--to--NIR non--thermal 
continuum emission and polarization, strong compact flat spectrum 
radio emission and superluminal motion. Similar properties are observed also  
in flat spectrum radio quasars (FSRQ) and the two types of active nuclei are 
often grouped together in the blazar class. 
From the spectroscopical point of view BL Lacs are  
characterized by quasi featureless optical spectra.
In fact their spectra are often dominated by the non--thermal 
continuum that arises from the nucleus. 
To this emission it is superimposed a thermal contribution 
due to the stellar component of the host galaxy.  
Like in other AGN, emission lines  could be generated by  
fluorescence in clouds surrounding the central 
black hole. Moreover, as it happens for high
z quasars in some cases absorption lines due to intervening gas in the halo 
of foreground galaxies can be observed in the spectra of BL Lacs 
and one can derive a lower limit to the redshift of the object.
The detectability of spectral features depends on the brightness of the 
nuclear source: in fact during low brightness states, intrinsic 
absorption features can be more easily  revealed, while during high 
states one can better discover intervening absorption systems.
Because of the strong contribution from the continuum 
the equivalent width (EW) of all these spectral features is often very 
small and their detection represents a challenging task. 

In the past decade a number of projects were carried out to derive the 
redshift of BL Lac objects. Most of these works were based on optical 
spectra collected with 4 m class telescopes, and are therefore limited by 
relatively low signal-to-noise  ratio (S/N), low spectral 
resolution and limited wavelength range 
\citep[e.g.][]{falomo93, stickel93, veron93, bade94,falomo94, falomo96, 
marcha96, drinkwater97,rgb, landt01, rector01, londish02, carangelo03,
hook03}. Recently, however, some observations with 8 m class telescopes were 
carried out \citep{heidt04,sowards05}.
Despite these efforts, a significant fraction of known BL Lacs (e. g. 50 \% in 
\citet{veron03} catalogue) have still unknown redshift.

In order to improve the knowledge of the redshift of BL 
Lacs we  carried out a project to obtain optical spectra of sources with still 
unknown or uncertain redshift using the European Southern 
Observatory (ESO) 8-m Very Large Telescopes (VLT). 
This allows one to 
improve significantly the S/N of the spectra and therefore the capability to 
detect faint spectral features. 
A first report on this work, giving the redshift of 12 objects, has been 
presented by \citet[][Paper I]{sbarufatti05a}, 
and here we refer on the results for the full sample of 42 observed sources. 

The outline of this paper is the following.
In section  \ref{sec:sample} we give some characterization of the 42 observed 
objects. The observation and analysis procedures are described in section 
\ref{sec:obsred}. 
In sections \ref{sec:results} and \ref{sec:notes} we report the results of 
our spectroscopic study. 
Finally in section \ref{sec:discussion} a summary and conclusions of this 
study are given. Throughout this paper we adopted the following cosmological 
parameters: H$_0$= 70 km s$^{-1}$ Mpc$^{-1}$, $\Omega_{\Lambda}$=0.7, 
$\Omega_m$=0.3.

\section{The sample}\label{sec:sample}

The sample of BL Lac objects (and candidates) observed with the VLT 
telescopes was selected from two extended lists of BL Lacs: the \citet{PG95} 
collection of BL Lacs  and the Sedentary Survey 
\citep[][in the following addressed as SS]{giommi99,giommi05}. 
The \citet{PG95} list contains all
objects identified as  BL Lacs belonging to the complete samples existing at 
the time of its compilation, selected in the radio, optical and X-ray bands 
(e.g.: 1 Jansky survey -- 1-Jy, \citet{stickel91}, 
Palomar-Green survey -- PG, 
\citet{green86}, Extended Medium Sensitivity Survey -- EMSS, 
\citet{gioia90}, Slew survey, \citet{perlman96}, White-Giommi-Angelini
catalogue -- WGA \citet{white94}). It 
includes also sources from the  \citet{hewitt93} and \citet{veron93} 
catalogues (in the latter case we checked that the source was still
included in the 2001 version), for a total of 233 objects. The criteria used 
to define a BL Lac object in \citet{PG95} depend on the sample of 
origin.
 In most cases, the 
EW of the lines is required to be $\leq$5 \AA, but also UV excess, 
optical polarization and variability, radio-to-optical spectral index are used 
as selecting criteria.
The SS was obtained cross-correlating the National Radio Astronomy 
Observatory (NRAO) Very Large Array (VLA) Sky Survey 
(NVSS) data \citep{nvss} with the ROSAT All Sky Survey--Bright Source
Catalogue (RBSC) list of sources \citep{voges99}. SS selected a 
complete sample of 150 High energy peaked BL Lacs 
\citep[HBL, see][for definition]{padovani95} down to a 3.5 mJy radio flux 
limit. BL Lac classification in the SS is based on the position of the sources 
on the $\alpha_{OX}-\alpha_{RO}$ plane.

\begin{figure}[htbp]
    \centering
\includegraphics[scale=.6]{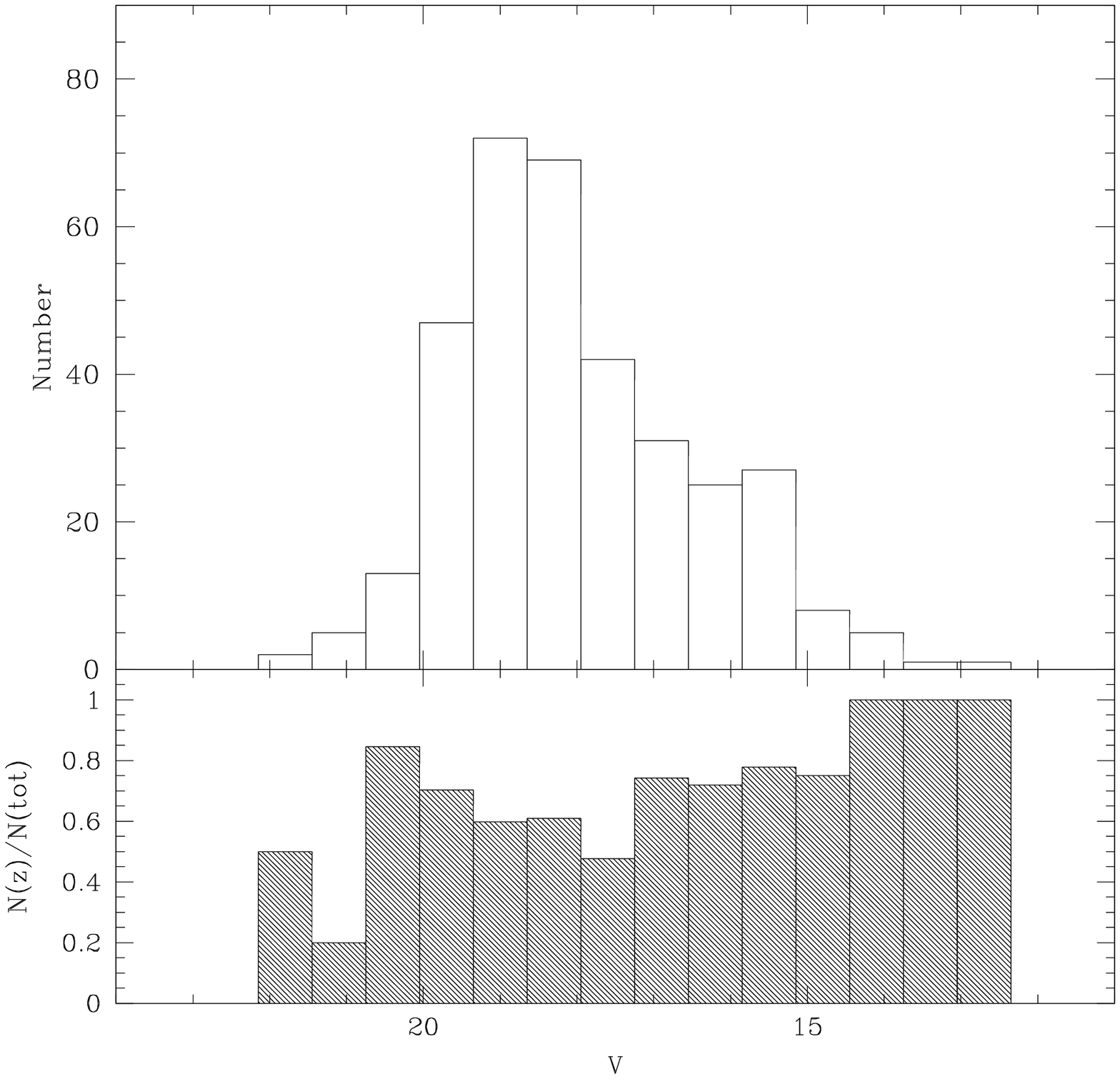}
  \caption{Upper panel: V magnitude distribution of BL Lac objects from 
  \citet{PG95} and SS.
  Lower panel: fraction of objects of known redshift as a function of the 
  magnitude.  }
\label{fig:distv}
\end{figure}

The  \citet{PG95} and SS datasets lead to a combined list  containing 
348 objects. The distribution of the V magnitude for these objects 
is reported in Fig. \ref{fig:distv}. The bulk of them have V  between 
15 and 20, and the the fraction of objects with unknown redshift 
increases with the apparent magnitude and reaches $\sim$ 50\% at the 
faintest magnitudes. 
Note, however, that also  at V $\sim$ 15-17 about 20\% of the sources 
have not known redshift. 
The total number of objects with unknown redshifts is 105.

From the combined list we selected sources with 
$\delta<$+15$^\circ$, for observability from the VLT site. 
Moreover to grant a sufficiently high S/N level of the optical spectra 
we required  V$<$22. Thus we gathered a list of 59 objects.
During three
observational campaigns, performed in service mode,  we completed this
optical spectroscopy program,
obtaining data for $\sim$70\% of the sample (42 sources). Our sample is
similar to the parent sample of 348 objects in terms of mean apparent
magnitude and subdivision in Low (LBL) and High energy peaked BL Lacs.

\section{Observations and data analysis}\label{sec:obsred}

Optical spectra were collected in service mode with the FOcal
Reducer and low dispersion Spectrograph \citep[FORS1,][]{fors} on the
VLT. The observations were obtained from April 2003 to 
March 2004 with UT1 (Antu) and from April to October 
2004 with UT2 (Kueyen).
We used the 300V+I grism combined with a 2'' slit, yielding a dispersion of 
110 \AA/mm (corresponding to 2.64 \AA/pixel) and a spectral resolution of 
15--20 \AA \ covering the 3800$-$8000 \AA \ range. The seeing during 
observations was in the range 0.5$-$2.5'', with an average of $\sim$1''.
Relevant informations on the sample objects are given in 
Table \ref{tab:results}.

Data reduction was performed using IRAF\footnote{IRAF (Image Reduction and
Analysis Facility) is distributed by the National Optical Astronomy 
Observatories, which are operated by the Association of Universities for 
Research in Astronomy, Inc., under cooperative agreement with the National 
Science Foundation.}\citep{tody86,tody93} following 
standard procedures for spectral analysis. This includes bias subtraction, 
flat fielding and  cleaning for bad pixels. For each target we obtained three 
spectra in order to get a good correction of cosmic rays and to check the reality
of weak features. The individual frames were then combined into a single 
average image. Wavelength calibration was performed using the spectra of a
Helium/Neon/Argon lamp obtained during the same observing night, 
reaching an accuracy of $\sim$ 3 \AA (rms). From these images we extracted 
one-dimensional spectra adopting an optimal extraction algorithm 
\citep{horne86} to improve the S/N. 

Although this program did not require optimal photometric conditions, 
most of the observations were obtained with clear sky. This enables us to 
perform a spectrophotometric calibration of the acquired data using 
standard stars \citep{oke90} observed in the  same nights. 
From the database of sky conditions at Paranal we estimate that a photometric 
accuracy of 10\% was reached during our observing nights. 
The spectra were  also corrected for Galactic extinction, using the 
law by \citet{cardelli89} and assuming values of E$_{B-V}$ from 
\citet{schlegel98}. 

\section{Results}\label{sec:results}

In Fig. \ref{fig:spec} we give the optical spectrum of each source. 
In order to show more clearly the continuum shape and 
the faint 
features we report both the flux 
calibrated and the normalized spectrum for each object. 
The main emission and absorption features are identified. 
Those due to the galactic interstellar gas are indicated 
as ``ISM'' and ``DIB'' (Diffuse Interstellar Bands, see section 
\ref{sec:resb}),  while telluric absorptions are marked as $\oplus$.

\subsection{The continuum emission}\label{sec:gfit}

 In a first approximation, the optical continuum of a BL Lac object is
due to the superposition of two components: the non-thermal emission of the 
active nucleus, Doppler-enhanced because of the alignment of the jet with the
line of sight, and the emission of the host galaxy. Depending on the
relative strength of the nucleus with respect to the galaxy light, the
spectral signature of the latter can be either easily detected or 
diluted beyond the point of recognition. Taking into account the
robust evidence that the host galaxies are giant ellipticals
\citep[e.g.][]{urry00}, to describe the continuum and derive the optical 
spectral index of the non-thermal component, we fitted a power law
($F_{\lambda}\propto\lambda^{-\alpha}$, the spectral indices are given
in Table \ref{tab:results}) plus the spectrum of a typical
elliptical galaxy as described by the \citet{kinney96} template. While
in most cases the contribution of the host galaxy was negligible, in 6 sources
it was not, and the luminosity of the host can thus be derived. For these six 
sources (three of them were presented in Paper I) 
we give the best fit decomposition in Fig.\ref{fig:gfit2} and report
 the parameters in 
Table \ref{tab:gfit}. The derived absolute magnitudes of the host galaxies
are consistent with the distribution of M$_R$ of BL Lac hosts given by 
\citet{sbarufatti05b}.

\subsection{Spectral features and redshifts}

The detection and the measurement of very weak spectral features is
difficult to assess because it depends on the choice of the
parameters used to define the spectral line and the continuum.  
In order to apply an 
objective method for any given spectrum we evaluate the minimum
measurable equivalent width (\EWmin) defined as twice the rms of the
distribution of all EW values measured dividing the normalized
spectrum into 30 \AA\ wide bins (details for this automatic
routine are given in Paper I).
We checked that the S/N ratio dependence inside the considered 
spectral range varies at most by 20 \%, remaining $<$10\% over a large 
wavelength range. This reflects  into a similar variation of \EWmin.
The  procedure for calculating \EWmin \ was applied to all featureless or 
quasi-featureless spectra to find faint spectral lines. All features above the 
\EWmin \ threshold, ranging from $\sim$ 1 \AA \ to 0.1 \AA \ in our data, were
considered as line candidates and were carefully visually inspected and 
measured. The results  are summarized in Tab. \ref{tab:results}. Based on the 
detected lines and the shape of the continuum we confirm the BL Lac 
classification for 36 objects, while 6 sources were reclassified.
Depending on the observed spectral properties the objects can be assembled in 
three groups. 

\subsubsection{Confirmed BL Lacs with measured z.}\label{sec:resa}

Twelve objects belonging to this group were reported  in paper I. 
Six more 
are presented here (Table  \ref{tab:results}). Three have 
redshift derived from emission lines (0723--008, z=0.128; 2131--021, z=1.284;
2223--114, z=0.997) and three from absorption  lines (1212+078, z=0.137;
1248--296, z=0.382; 2214--313, z=0.460). Details on each 
source are given in section \ref{sec:notes}.

\subsubsection{ Misclassified and uncertain nature objects.}\label{sec:resc}

Despite their classification as BL Lac
objects in one or more input catalogues, six sources have spectra
incompatible with this identification. Five of them were reclassified either 
as quasars (0420+022, 1320+084) or stars (1210+121, 1222+102, 1319+019), while
object 0841+129 remains of uncertain nature.

\subsubsection{Lineless BL Lacs.}\label{sec:resb}

In spite of the high S/N 18 objects exhibit spectra lacking any intrinsic 
feature. In several spectra we clearly see absorption features from the 
interstellar medium (ISM) of our Galaxy. In particular, we are able to detect 
CaII $\lambda\lambda$3934,3968, NaI $\lambda$5892 atomic lines, and a number 
of DIBs $\lambda\lambda$ 4428,4726,4882,5772, generated by complex molecules 
in the ISM \citep[e.g.][and references therein]{galazutdinov00}.
In Fig. \ref{fig:ISM} we report the average spectrum of the interstellar 
absorptions. 
In three cases absorption lines from intervening gas are detected, leading to
lower limits on the redshift of the objects (0841+129, 
z$>$2.48; 2133--449, z$>$0.52; 2233--148, z$>$0.49).

For these 18 sources we have estimated a redshift lower limit based on the
\EWmin \ of their spectra and the apparent magnitudes of the nuclei. We 
report \textbf{these} in Table \ref{tab:results}. The procedure to obtain 
these limits is described in section \ref{sec:ew2z}.

\subsubsection{Redshift lower limits procedure.}\label{sec:ew2z}

In this section we describe the procedure to obtain redshift lower 
limits for BL Lacs with lineless spectra (see Table \ref{tab:results}) from 
the \EWmin \ of the spectrum and the observed magnitude of the object. Under 
the assumption that the host galaxy luminosity is confined in a narrow range 
\citep{sbarufatti05b} it is in fact possible to constrain the position of the 
source on the nucleus-to-host flux ratio ($\rho$) \textit{vs} redshift plane.

We assume that the observed spectrum of a BL Lac object is given 
by the contribution of two components: 1- a non-thermal emission from the 
nucleus that can be  described by a power law 
($F(\lambda)= C  \lambda ^{ -\alpha}$, where $C$ is the 
normalization constant); 2 - a thermal component due to the host galaxy. 
Depending on the relative contribution of the two components the optical 
spectrum will be dominated by the non-thermal (featureless) emission or by 
the spectral signature of the 
host galaxy. The observed equivalent width (EW$_{obs}$) of a given spectral 
absorption line is diluted depending on the ratio of the two 
components. Detection of this spectral feature requires a spectrum with
a sufficiently high S/N. This is illustrated in  Fig 
\ref{fig:specsim}, where a simulated spectrum ($\rho$=5, z=0.5) is reproduced
with two different S/N ratios. The S/N=300 spectrum grants a secure detection
of the CaII features, while with S/N=30 the lines are undetected.

In order to estimate the redshift of an object from the \EWmin \ 
we need to know the relation between EW$_{obs}$ and the nucleus-to-host flux
ratio $\rho$. For a spectral absorption line of intrinsic equivalent width
EW$_0$ the observed equivalent width is given by the  relation 
\citep[see also][]{phd}:

\begin{equation}\label{eq:1}
\mathrm{EW}_{obs}=\frac{(1+z) \times \mathrm{EW}_0}{1+\rho \times A(z)}
\end{equation}

The nucleus-to host ratio $\rho$ can be represented by
 
\begin{equation}\label{eq:2}
\rho(\lambda)=\frac{F(\lambda)}{G(\lambda)}
\end{equation}

where $G(\lambda)$ is the giant elliptical spectral template by
  \citet[][see also section \ref{sec:gfit}]{kinney96}, and A($z$) is a
correction term that takes into account the loss of light inside the
observed aperture. In this work the aperture is a 2''$\times$6'' slit that
captures $\gtrsim$~90\% of the nuclear light, but not the whole surrounding
galaxy that is more extended than the aperture (in particular for low z
targets). In order to estimate this effect we evaluated the amount of light
lost from the galaxy through the aperture in use from 
simulated images of BL Lacs (point source plus the host galaxy). The main 
parameters involved are the shape and the size of the host galaxy. According 
to the most extensive imaging studies of BLL \citep{falomo96,wurtz96,falomo99,
heidt99,nilsson03,urry00} we assumed that the host galaxy is a giant 
elliptical of effective radius R$_e$= 10 kpc. The fraction of starlight lost 
then depends on the redshift of the object and is particularly significant 
at z$<$ 0.2, producing the bending of the curves in Fig. \ref{fig:apeffect}.

Since we want to refer the observed equivalent width to the 
nucleus-to-host ratio $\rho_0=\rho(\lambda_0)$ at a fixed wavelength 
$\lambda_0$, equation (\ref{eq:1}) can be rewritten as:
\begin{equation}\label{eq:ewnh}
\mathrm{EW}_{obs}=\frac{(1+z) \times \mathrm{EW}_0}{1+\rho_0 \times \Delta \times A(z)}
\end{equation} 
where $\Delta(\lambda)$ is the nucleus-to-host ratio normalized to that at 
$\lambda_0$ ($\Delta(\lambda)=\rho(\lambda)/\rho(\lambda_0)$; see Fig. \ref{fig:delta}).

On the other hand the quantity  $\rho_0$ depends also on the observed
 magnitudes of the object, since
\begin{equation}\label{eq:magnh}
\log(\rho_0) = -0.4  [\mathrm{M}_n(z) - \mathrm{M}_h(z)] 
\end{equation} 
where M$_n$ is the nucleus absolute magnitude and M$_h$ is the host 
absolute magnitude, and
\begin{equation}
\mathrm{M}_n(z)= \mathrm{m}_n +5 -5 \log{\mathrm{d}_l(z)} - \mathrm{k}_n(z)
\end{equation}
where m$_n$ is the nucleus apparent magnitude, d$_L$($z$) is the 
luminosity distance and k$_n$($z$) is the nucleus k-correction, 
computed following \citet{wisotzki00}. The absolute magnitude of the host is
\begin{equation}
\mathrm{M}_h(z)= \mathrm{M}_h^* - E(z)
\end{equation}
where M$_h^*$=--22.9 is the average R band magnitude of BL Lac hosts at z=0 
and E($z$) is the evolution correction, as given by \citet{bressan98}.

An example of the procedure described above is given in
Fig. \ref{fig:ew2z}, where the relationships between log($\rho_0$) and the
redshift for a given value of \EWmin \ and m$_n$ are shown.
The intersection of the two curves yields a lower limit to the redshift of the 
target. When it goes beyond the observed spectral range, we set the redshift 
limit to the value corresponding to the considered feature reaching the upper 
limit of the observed wavelength range (z$\sim$1 in the case of CaII 
$\lambda$3934 line). The uncertainty of this procedure depends
mainly on the spread of the distribution of the host galaxy
luminosity. This issue is discussed in \citet{urry00} and in 
\citet{sbarufatti05b}, where it is shown that the 64 BL Lacs hosts of known 
redshift resolved with HST are well represented by an elliptical of 
M$_R$=-22.9, with 68\% of them in the interval -23.4 -- -22.4. 

This procedure can be used for any absorption line belonging to the host 
galaxy and for which an estimate of the un-diluted EW is available. 
In this work we considered the  
CaII absorption line at $\lambda$=3934 \AA (EW$_0$=16 \AA), we assumed
a power law spectral index $\alpha$=0.7 \citep{falomo93}, and we referred 
to the effective wavelength of the R band ($\lambda_0$=6750 \AA) to compute 
$\rho_0$ (which implies $\Delta$=4.3).

In order to test this procedure we considered eight BL Lacs for which the CaII 
line of the host galaxy has been measured. 
Five of these objects derive from the observations discussed here 
and in paper I, three others are from observations obtained at the ESO 3.6 
\citep{carangelo03, phd}. These spectra are reported in Fig. 
\ref{fig:zestspec} and the relevant parameters are given in Table 
\ref{tab:resultscomp}. The comparison between the redshifts estimated 
by our procedure with the spectroscopic ones indicates a reasonable good 
agreement (see Fig. \ref{fig:zestcomp}).

\section{Notes to individual objects.}\label{sec:notes}

\paragraph{0047+023} 
This compact and flat spectrum radio source was classified as a BL Lac
by \citet{hewitt93} on the basis of UV color and featureless spectra.
Further featureless optical spectra obtained by \citet{allington91,
veron93} confirmed the BL Lac.  Even in our S/N$\sim$ 80 spectrum no
spectral features were found.  Based on the minimum detectable EW the
source is most likely at z $>$ 0.82.

\paragraph{0048--097} Previous optical observations  of this well known  
BL Lac object  belonging to the 1-Jy sample, reported a featureless 
spectrum \citep{stickel91, falomo94}. \citet{rector01}, however, 
suggested the presence of an emission line at 6092 \AA \ (possibly identified 
with [OII] $\lambda$3727 at z=0.634 or [OII] $\lambda$5007 at z=0.216). 
\citet{falomo96} proposed z$>$0.5, based on the non detection of the host 
galaxy in the optical images of the source. Our S/N=250 optical spectrum does 
not confirm the presence of the emission line at 6092 \AA, and 
apart of some telluric lines  and a number of Galactic 
absorptions it is found featureless.
 From our \EWmin \ estimate, we  infer that this source is at z$>$0.3. 

\paragraph{0420+022} \citet{fricke83} classified this source as a BL Lac 
candidate on the basis of a featureless (although noisy) optical 
spectrum. \citet{ellison01} through an unpublished optical spectrum 
propose a redshift z=2.277 and  classify the source as a radio loud QSO. 
In our optical spectrum we are able to clearly detect 
emission lines  Ly$_{\alpha}~\lambda$1419  OVI] $\lambda$1034, 
CIV $\lambda$1549 and CIII] $\lambda$1909, at z = 2.278. 
A recent spectrum obtained by \citet{hook03}  also confirm our findings 
and the  classification of the object as a QSO. 

\paragraph{0422+004} This object is a well known radio selected BL Lac, 
included in the \citet{hewitt93} catalogue. \citet{falomo96} detected the host 
galaxy with ground-based imaging, proposing z$\sim$0.2--0.3. The 
optical spectrum taken by \citet{falomo94} is featureless. Our spectrum 
(S/N=230) does not show evidence for intrinsic spectral features 
from the host, suggesting a very high N/H ratio. Interstellar absorptions 
from NaI $\lambda$5892 and DIBs at 5772 and 4726 \AA \ are well detected. 
Based on \EWmin, we estimate z$>$0.31.

\paragraph{0627--199} 
\citet{hook03} obtained a lineless spectrum for
this radio selected BL Lac object. Our VLT spectrum, of moderate
S/N (50), shows no spectral features. From \EWmin we set 
z$>$0.63.

\paragraph{0723--008} \citet{wills76} classified this source as a Narrow 
Emission Line Radio Galaxy based on an optical spectrum, 
giving z=0.127. \citet{rusk85} report an
optical polarization of 1.5 \%, classifying the source as a Low
Polarization QSO. \citet{VV01} report this source as a BL Lac object.  
\citet{henriksen84} gives broad band indices $\alpha_{RO}$=0.7 and 
$\alpha_{OX}$=1.0, which are compatible with a BL Lac or a FSRQ classification.  
Our optical spectrum is clearly dominated by a non thermal emission with 
spectral index $\alpha$=0.7. Superposed to this, strong narrow emission lines 
and absorption lines from the underlying host galaxy at z=0.127 are
detected, confirming the redshift.  From
the values of the spectral indices and the measured EW for the
spectral lines, we suggest that this object is of intermediate nature
between a BL Lac and a quasar.

\paragraph{0841+129} This source,
first identified by C. Hazard \citep[see][and references therein]{jaunsen95},
is a Damped Lyman $\alpha$
Absorption (DLA) QSO at z$>$2.48 as derived from
the two very strong DLAs at $\sim$4100 and $\sim$4225 \AA \ 
\citep[see for example][and references
therein]{pettini97,prochaska01,warren01}.
The  classification as a BL Lac object was motivated by the absence of
prominent emission
lines  \citep{hewitt93}.

Our spectrum, in addition to several absorption lines, exhibit
three possible broad emission structures at$\sim$4310, $\sim$4850 and
$\sim$5370 \AA.
These could be interpreted as NV $\lambda$1240, SiIV $\lambda$1397 and 
CIV $\lambda$1549 at
z$\sim$2.47.
This is consistent with  z$\sim$2.5, deduced
from the observed position of the onset of the absorption of the
Ly$_{\alpha}$ forest \citet{warren01}.
An alternative explanation is, however, that these  structures are
pseudo-emissions resulting from the depression of the continuum caused by the 
envelope of many unresolved narrow absorption features.
Higher resolution spectra of the object in the spectral range 4200 to 5800 \AA
\  are needed to distinguish between the two possibilities.

\paragraph{1210+121} \citet{hazard77} proposed that this object was
  the optical counterpart of a radio source in the Molongo Catalogue
\citep[MC2][]{sutton74}; the separation was however 16''.
\citet{zotov79} reported large optical variability 
and polarization, apparently reinforcing the identification. \citet{baldwin73}
found a featureless  optical spectrum. Our VLT spectrum clearly 
shows that the source is a type B star in our Galaxy.

\paragraph{1212+078} Our VLT spectrum clearly shows the presence of a strong 
thermal component due to the host. We  detected CaII
$\lambda\lambda$3934, 3968, G band $\lambda$4305, MgI $\lambda$5175 and
H$_{\alpha}~\lambda$6563 in emission at z=0.137, confirming the redshift
estimated by \citet{perlman96}.  The contribution of the non-thermal
component is visible in the bluest part of the spectrum.  The best fit
decomposition of the spectrum gives $\alpha$=1.17 for the non-thermal
component and M$_R$=-22.0 for the host. Though this is somewhat fainter than
expected for a BL Lac host galaxy, we can not exclude that
given the low redshift and the consequent large apparent size of the
host, part of the light did not enter in the slit.

\paragraph{1222+102} 
This is a blue stellar object in the direction of
the Virgo-Coma cluster. Its apparent position in the sky is very close
to the center of the galaxy NGC 4380, still well inside the galaxy 
boundaries.  The projected separation to the nucleus at the redshift of the 
galaxy is $\sim$10 kpc. The object is considered a candidate BL Lac 
in the \citet{burbidge87} list, selected because of its UV excess.
\citet{arp77} reports the observation of a featureless
spectrum. \citet{burbidge96} estimates this object a
possible candidate of expulsion from a galactic nucleus.
The sharp absorption lines detected in our spectrum clearly indicate 
a stellar origin. The measured colors lead to a temperature of $\sim$10000 K. 
If the object were a main sequence or a supergiant star, the corresponding
distance will put it outside the Galaxy, but not at the distance of NGC 4380. 
We are therefore led to consider a white dwarf, which would be at 100--200
pc. The absence of H lines indicates a DQ or DXP white dwarf \citep{schmidt01,schmidt03}. 
Some of the lines are referable to HeI and CI transitions. The object clearly 
deserves further study; in particular polarization measurements would 
be interesting.

\paragraph{1248--296} \citet{perlman96} obtained a low S/N spectrum of this
source, and proposed a BL Lac at z=0.487 based on the possible
detection of the host galaxy features. In our VLT spectrum CaII, G
band, H$_{\beta}$ are clearly detected at z=0.382, confirming the
findings of \citet{woo05}, while in the blue part the contribution of
a non-thermal component is clearly visible.  The best fit
decomposition gives $\alpha$=0.92 for the non-thermal component
visible below 5000 \AA, and M$_R$=-22.7 for the host, in good
agreement with result from the direct detection of the host in HST imaging
\citep{urry00,sbarufatti05b}.

\paragraph{1319+019} This object was initially selected 
as a  BL Lac candidate on the basis of the University of Michigan 
objective prism survey \citep[][]{macalphine81} designed to find AGN 
and it is included as BL Lac in the \citet{VV01} catalogue. 
No radio counterpart for this
source has been found in literature. Later \citet{thompson90a} proposed its
classification as a BL Lac, based on a low S/N optical spectrum that
was found featureless.  In our much better quality spectrum we clearly
see many absorption features that characterize the object as a
galactic star of spectral type $\sim$A.  Our findings are also in agreement 
with the spectral classification of the 2dF QSO Redshift survey 
\citep[2QZ, see][]{croom04}.

\paragraph{1320+084} This source is part of the BL Lac sample extracted 
from the EINSTEIN Slew Survey and a radio counterpart was reported by
\cite{perlman96}.  Our VLT data show the source has a QSO like
spectrum at z=1.5, in contrast with a featureless spectrum observed
by \cite{perlman96}.  Several intervening absorption lines, in
particular MgII at z=1.347 were also
detected.

\paragraph{1349--439} The spectrum of this X-ray selected BL Lac 
\citep{dellaceca90}, shows a number of absorption lines
from the interstellar medium: CaII $\lambda\lambda$3934,3968, the 5772 \AA \
DIB, NaI $\lambda$5892. No intrinsic features were detected, and the deduced 
redshift lower limit is z$>$0.39. As already pointed
out by \citet{veron96}, the value z=0.05 sometime reported for this object
is consequence of a confusion with the nearby Seyfert 1 galaxy Q 1349-439. 

\paragraph{1442--032} This X-ray source, the radio counterpart of which was 
found in the NVSS survey, was first classified as a BL Lac in the
RBSC-NVSS sample by \citet{bauer00}, and then confirmed by the
SS. There are no published optical spectra for this source. Our
optical spectrum is featureless, with the exception of the NaI
$\lambda$5892 absorption feature from our galaxy ISM. The \EWmin \ value
for this objects leads to z$>$0.51.

\paragraph{1500-154} This X-ray selected BL Lac is part of the 
RSBC-NVSS sample \citep{bauer00} and enters in SS.  
No previous optical spectroscopy has been found in the literature. Our 
spectrum is completely featureless, leading to z$>$ 0.38 from the obtained 
\EWmin.

\paragraph{1553+113} This source is an optically selected BL Lac from the 
Palomar-Green survey.  The redshift estimate z=0.360 given in the
\citet{hewitt93} catalogue was disproved by later spectroscopy
\citep{falomo90,falomo94}.  While no intrinsic features were
detected in our S/N=250 VLT spectrum, a number of absorption lines due
to our galaxy ISM were revealed: CaII $\lambda\lambda$3934,3968, NaI
$\lambda$5892 and DIBs at 4428,4726,4882,5772 \AA. The \EWmin \ estimate
\ for this object gives a limit z$>$0.09.

\paragraph{1722+119}  \citet{griffiths89} reported 
a tentative redshift z=0.018 for this X-ray selected, highly 
polarized BL Lac. This estimate was not confirmed by more recent observations 
\citep{veron93,falomo93,falomo94}. Our VLT spectrum  (S/N=350) shows only 
absorption features due to our galaxy ISM: CaII $\lambda\lambda$3934,3968, 
NaI $\lambda$5892 and  DIBs at 4428 \AA, 4726 \AA, 4882 \AA, 5772 \AA, with no 
evidence of intrinsic features. 
From the minimum \EWmin \ we derive z$>$0.17.

\paragraph{2012--017} Consistently with previous observations of this radio selected 
BL Lac \citep{white88,veron90,falomo94}, also our S/N=130 VLT spectrum is
featureless. The optical spectral index is $\alpha$=0.49, 
in marginal agreement with $\alpha$=0.33$\pm$0.12
reported by \citet{falomo94}. From \EWmin \ we derive  z$>$0.94.

\paragraph{2128--254} The spectrum of this X-ray selected BL Lac 
candidate is reported as  featureless by SS. We confirm this result
and set a lower limit of z$>$0.86 for the redshift.

\paragraph{2131--021} \citet{rector01} and \citet{drinkwater97} proposed
a redshift of 1.285 for this source, based on the detection of CIII]
$\lambda$1909, MgII $\lambda$2798 and [OII] $\lambda$3727, opposed to the
z=0.557 suggested by \citet{wills76}. While [OII] falls outside our
spectral range, we  confirm the presence of CIII] and MgII emission lines at 
z=1.283, also detecting the fainter CII] $\lambda$2326 feature at 
the same redshift.

\paragraph{2133--449} This source was discovered because of its optical 
variability by \citet{hawkins91}. Optical spectroscopy by \citet{hawkins91}
and \citet{heidt04} led to completely featureless spectra. Our VLT
observations clearly show the presence of an intervening absorption feature 
at 4250 \AA, a tentative identification of which is intervening 
MgII at z=0.52 \citep[see][]{churchill05}. 
The lower limit on z derived 
from \EWmin \ is z$>$0.98 .

\paragraph{2136--428} The spectrum obtained by \citet{hawkins91}, who 
discovered this source studying its optical variability, is completely
featureless. Our VLT observations shows several absorption features
due to the ISM of our galaxy: DIB at 4428 \AA, 4726 \AA, 4882 \AA \
and 5772 \AA, CaII $\lambda\lambda$3934, 3968 and NaI$\lambda$ 5892
atomic lines.  The feature at 5942 \AA \ could be CaII $\lambda$3968
at z=0.497. Since at this redshift the CaII $\lambda$3934 should
fall at 5890 \AA, where it will be strongly contaminated by the
interstellar NaI absorption, the redshift estimate is only
tentative. The lower limit deduced by the minimum measurable EW is
z$>$0.24. 

\paragraph{2214--313} Our VLT spectrum of this object clearly shows 
the typical spectral signature of the host galaxy (CaII
$\lambda\lambda$3934,3968 and G band $\lambda$4305) at z=0.46.
The best fit decomposition gives $\alpha$=0.9 for the non-thermal component
and M$_R$=--22.3 for the host.
Previous optical spectroscopy performed by \citet{bade94} with the ESO
3.6m telescope failed to detect any spectral feature.

\paragraph{2223--114} Optical observations of this radio source obtained by 
\citet{veron93b} did not show any intrinsic spectral feature.  In our
spectrum, that extends further in the red, we detect a single narrow
emission line at $\lambda$7367 (EW = 5 \AA). This is a real feature
since it clearly appears on each of the 3 individual spectra
(see section 3).  A possible identification of this line is
[OII]$\lambda$ 3727 at z=0.977, while MgII $\lambda$2798 gives  
z=1.633. We discarded this second classification because the line 
FWHM (1200 km s$^{-1}$), is typical for a narrow line such [OII], 
while for MgII a larger value would be expected. Moreover, with a 
MgII identification both CIV $\lambda$1549 and CIII] $\lambda$1909  
broad lines would be expected inside the observed spectral range, but no other 
features are detected.

\paragraph{2233--148} The redshift z=0.325 reported by \citet{johnston95} 
is due to confusion with the source HB89 2233+134 in
\citet{schmidt83}.  \citet{drinkwater97} report an intervening system
at z=0.609, but without giving an identification of the corresponding
absorption feature. We detect several absorption features on the
spectrum. In particular we propose to identify the features at 4165
and 4183 \AA \ as MgII at z=0.492, while, using the \EWmin \ estimate
from the spectrum, z$>$0.65 is found.

\paragraph{2254--204} Previous optical spectroscopy \citep{veron93,hook03}
of this BL Lac object from the 1Jy sample showed completely featureless 
spectra. With VLT we are able to
detect faint interstellar absorptions of CaII $\lambda\lambda$3934, 3968 and
NaI$\lambda$ 5892, but no intrinsic or intervening spectral lines are found.
The inferred redshift limit is z$>$0.47.

\paragraph{2307--375} This source was first classified as a BL 
Lac in the RSBC-NVSS sample \citep{bauer00}. The classification was then 
confirmed by the SS. No previous optical spectroscopy has been 
published. Our VLT spectrum is featureless, allowing us to set only 
a lower limit to the redshift of z$>$1. 

\paragraph{2342--153} This source is part of the EMSS 
sample of BL Lac objects.  Our VLT data, as well as previous 
optical spectroscopy with the 6.5 m telescope of Multi Mirror 
Telescope Observatory \citep{rector00} showed a featureless
spectrum. From \EWmin \  we derive z$>$1.

\paragraph{2354--021}

This object was discussed in paper I. 
Here we report only the spectrum, in Fig. \ref{fig:spec}.

\paragraph{2354--175} This X-ray source from ROSAT All Sky 
Survey, is classified as BL Lac candidate in the RBSC-NVSS sample 
\citep{bauer00} 
and in the SS. No previous spectroscopy 
was published in literature. Our S/N=150 VLT spectrum is featureless, 
allowing only to set a lower limit of 
z$>$0.85 to the redshift.

\section{Summary and conclusions}\label{sec:discussion}

Out of 42 objects observed we confirm the BL Lac classification for 36
sources and for 18 of them we are able to measure/confirm the
redshift. This information allows us to derive 
the luminosity of the objects.  The distribution in the V band
luminosity-distance plane is indeed fully consistent with what
observed for BL Lacs of known redshift in the combination of the 
\citet{PG95} and the SS sample (see Fig. \ref{fig:distz}). 
We note that the sources in this combined list are affected by 
the typical selection effect of incomplete, flux limited 
samples: the 
envelope of the objects follows in fact the expected behavior for 
sources with constant V magnitude, centered around V=18, with a spread of 
$\sim$ 3 mag.
The objects discussed here (filled circles and lower limits) follow the same 
distribution, with absolute magnitudes ranging between -21.5 and -27.5,
slightly increasing with the redshift.

In 18 cases the optical spectra remain lineless in spite of the high  S/N 
of the obtained optical spectra. This indicates that if they are hosted by 
galaxies of standard luminosity they have likely very luminous or extremely 
beamed nuclei \citep[see also][]{sbarufatti05b}. 
In the latter case one may expect to see the most extreme cases of
relativistic beaming, making these sources ideal targets for
milli-arcsec resolution radio observations. 
Alternatively, if the host galaxies were under-luminous, these objects could be 
rare  examples of dwarf galaxies hosting an AGN \citep[see][]{sbarufatti05b}.  

The high S/N of most of the optical spectra obtained at VLT represents a 
frontier for the determination of the redshift of BL Lacs with current 
instrumentation and further improvement of the issue will not be easy 
to get. In particular the four brightest objects (R$<$16: 0048--099, 
1553+113, 1722+119, 2136-428) we observed at the VLT have values of \EWmin \ 
smaller than 0.25 \AA. These objects belong to an interesting sub-population 
of BL Lac objects with extreme nuclear (and/or host) properties for which 
it is actually not possible to derive the intrinsic 
physical parameters. 
One  possibility is to image  the source when it is 
particularly faint in order to improve the detection of  
the host galaxy and to derive  an imaging redshift \citep{sbarufatti05b}. 
Deep spectroscopic observations in the near-IR may also prove to be effective
in the determination of the redshift considering this region of the spectrum 
is poorly known. 

\acknowledgments
We would like to thank the referee, Dr. John Stocke, for his accurate and
helpful comments, which allowed us to improve the quality of this work.

\clearpage
\onecolumn


\begin{figure}[htbp]
    \centering
\includegraphics[scale=.7]{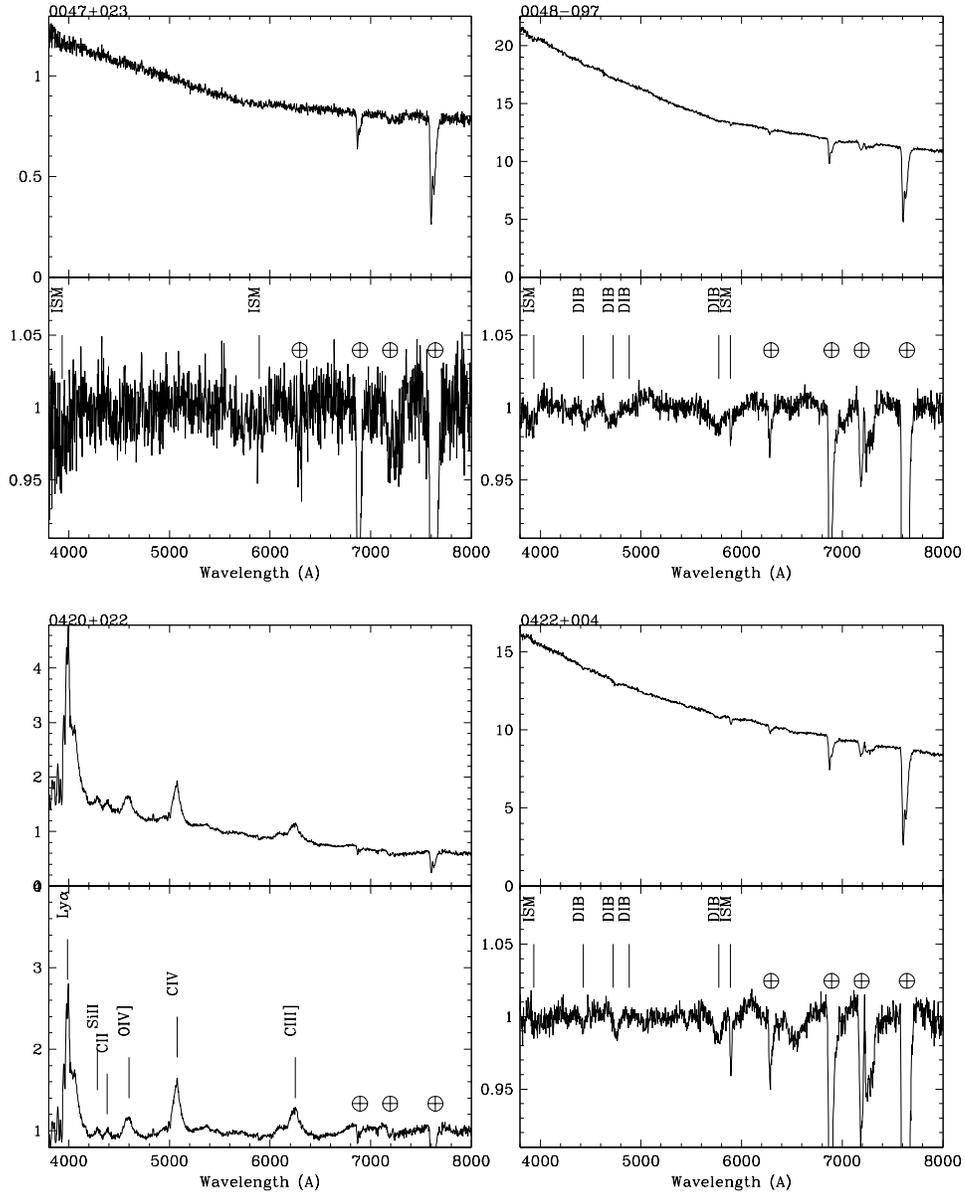}
  \caption{Spectra of the observed objects. Top panels: flux calibrated 
   dereddened spectra.
   Bottom panels: normalized spectra. 
   Telluric bands are indicated by $\oplus$, spectral lines are marked by the
   line identification,intervening MgII absorption systems are
   reported as "int. MgII", unidentified intervening systems are 
   indicated with *, absorption features from atomic species in the 
   interstellar medium of our galaxy are labeled by ISM, diffuse interstellar 
   bands by DIB.}
\label{fig:spec}
\end{figure}
\addtocounter{figure}{-1}

\begin{figure}[htbp]
    \centering
  \resizebox{\hsize}{!}{\includegraphics{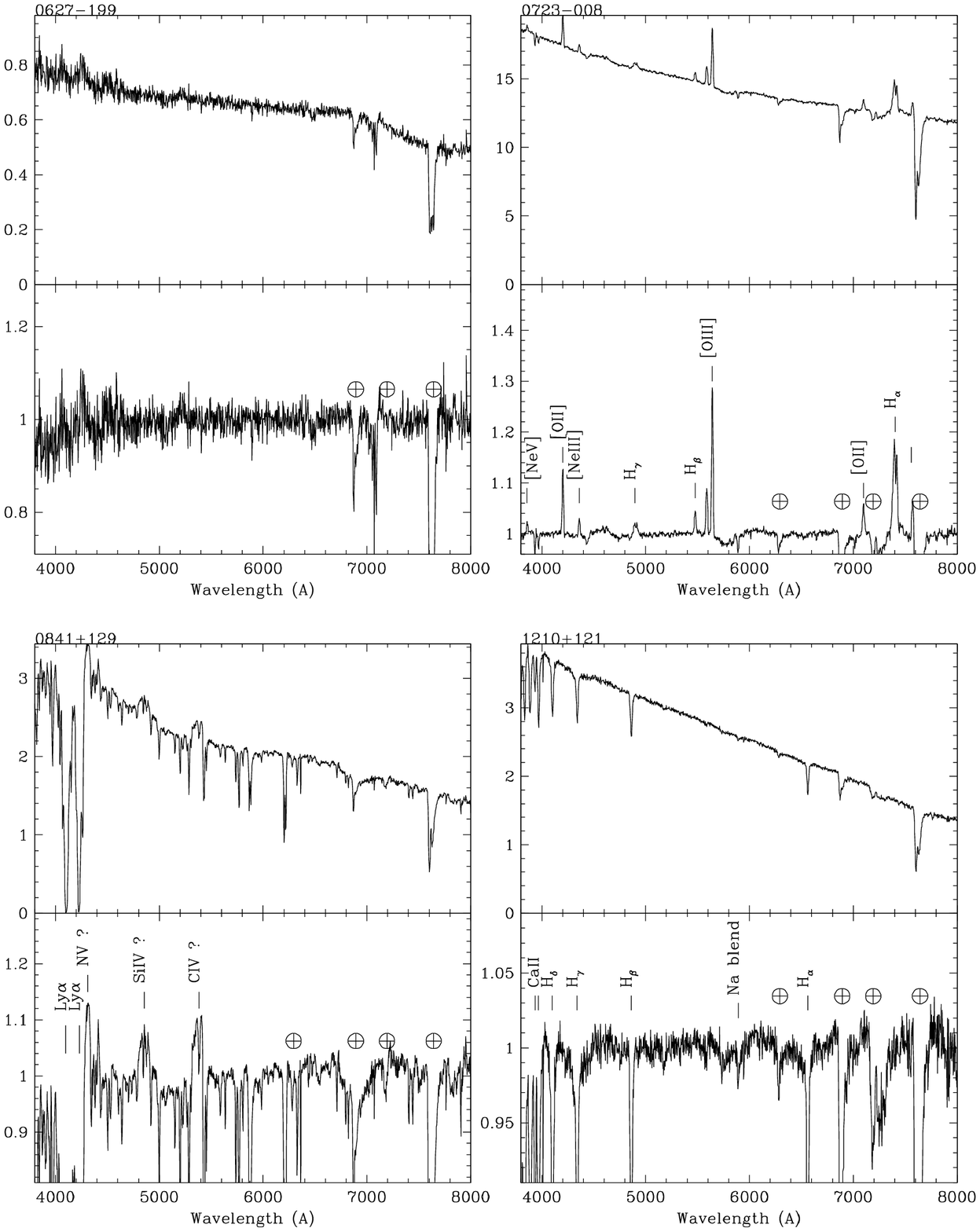}}
  \caption{continued}

\end{figure}
\addtocounter{figure}{-1}

\begin{figure}[htbp]
    \centering
  \resizebox{\hsize}{!}{\includegraphics{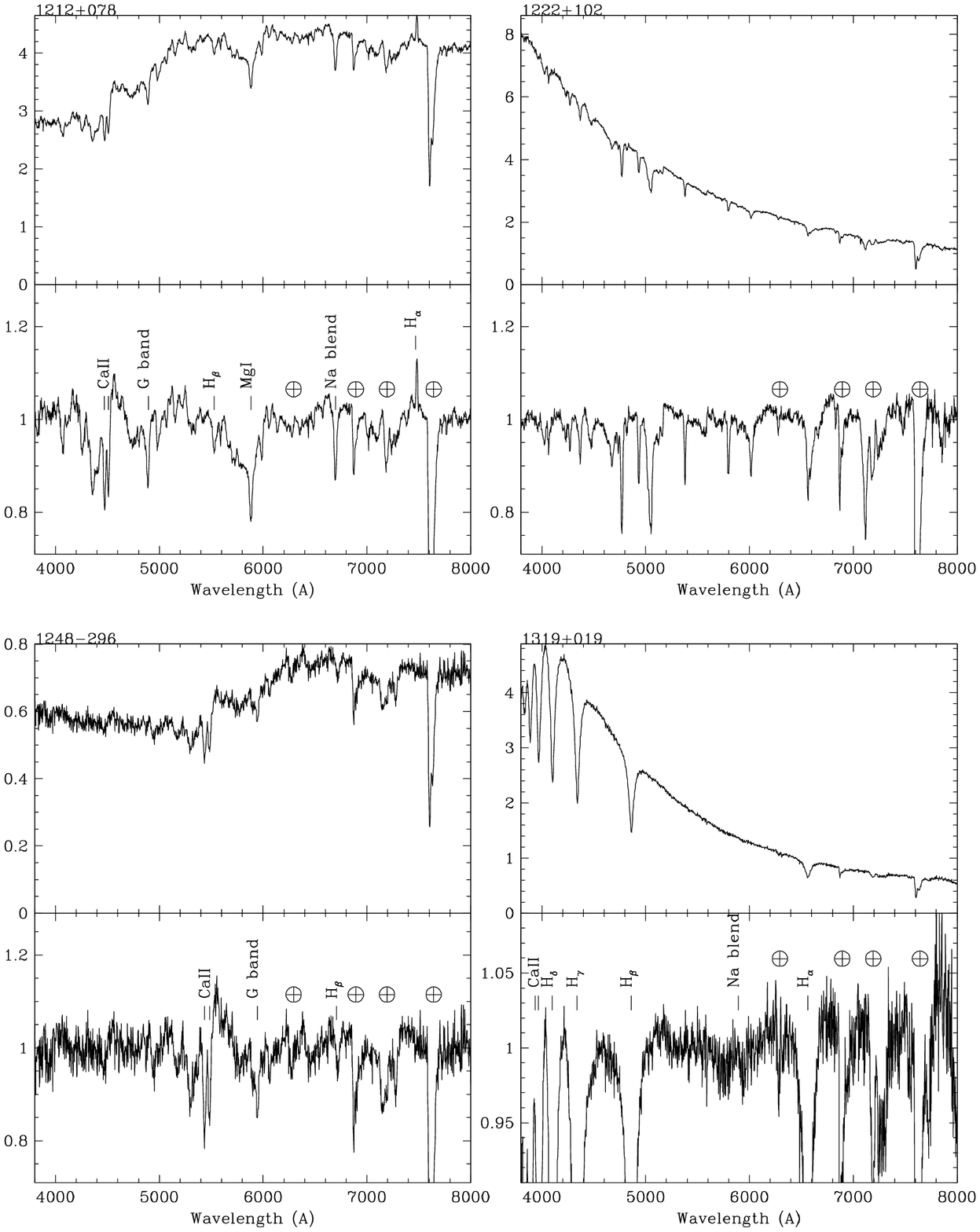}}
  \caption{continued}

\end{figure}
\addtocounter{figure}{-1}

\clearpage

\begin{figure}[htbp]
    \centering
  \resizebox{\hsize}{!}{\includegraphics{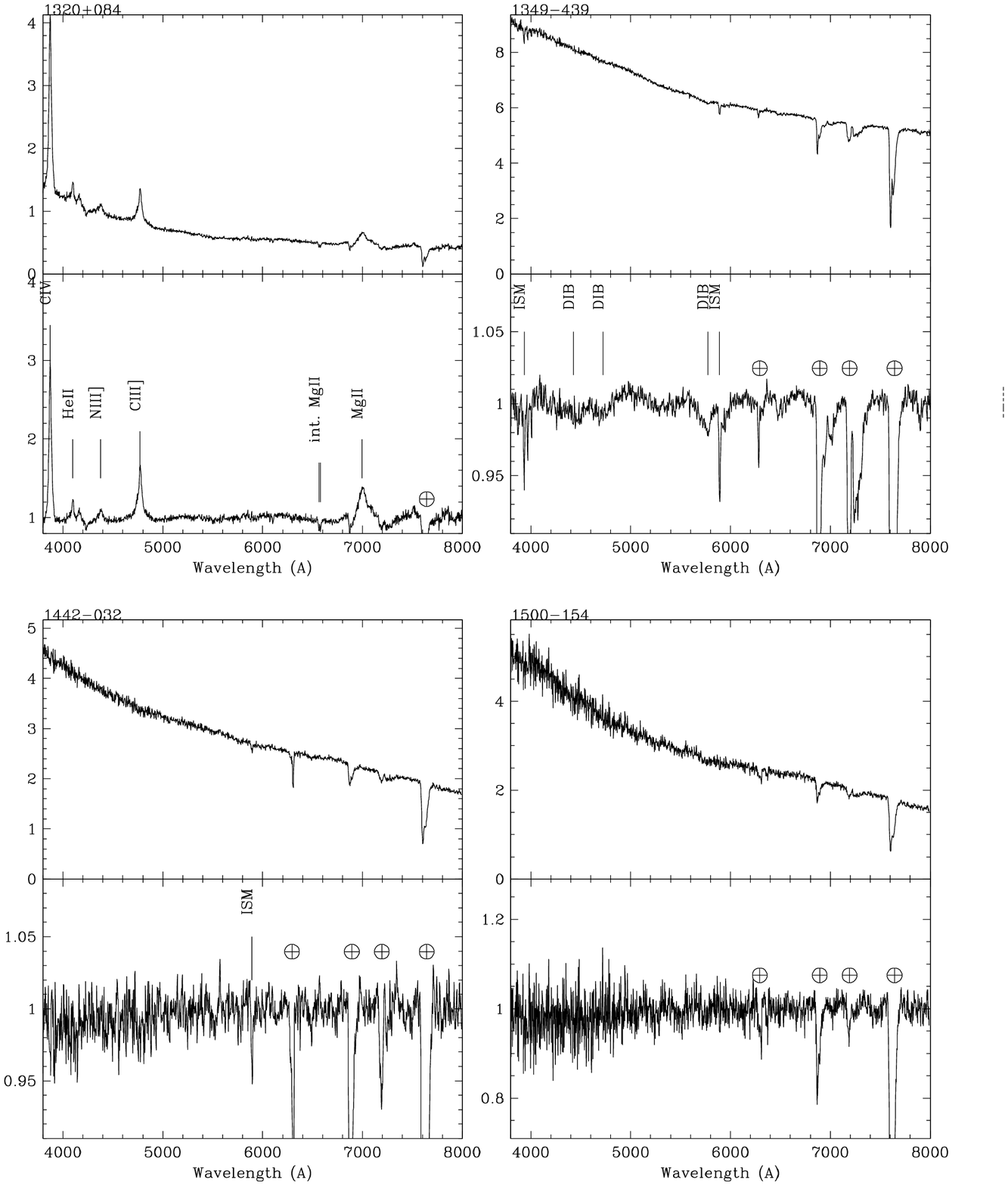}}
  \caption{continued}

\end{figure}
\addtocounter{figure}{-1}

\clearpage

\begin{figure}[htbp]
    \centering
  \resizebox{\hsize}{!}{\includegraphics{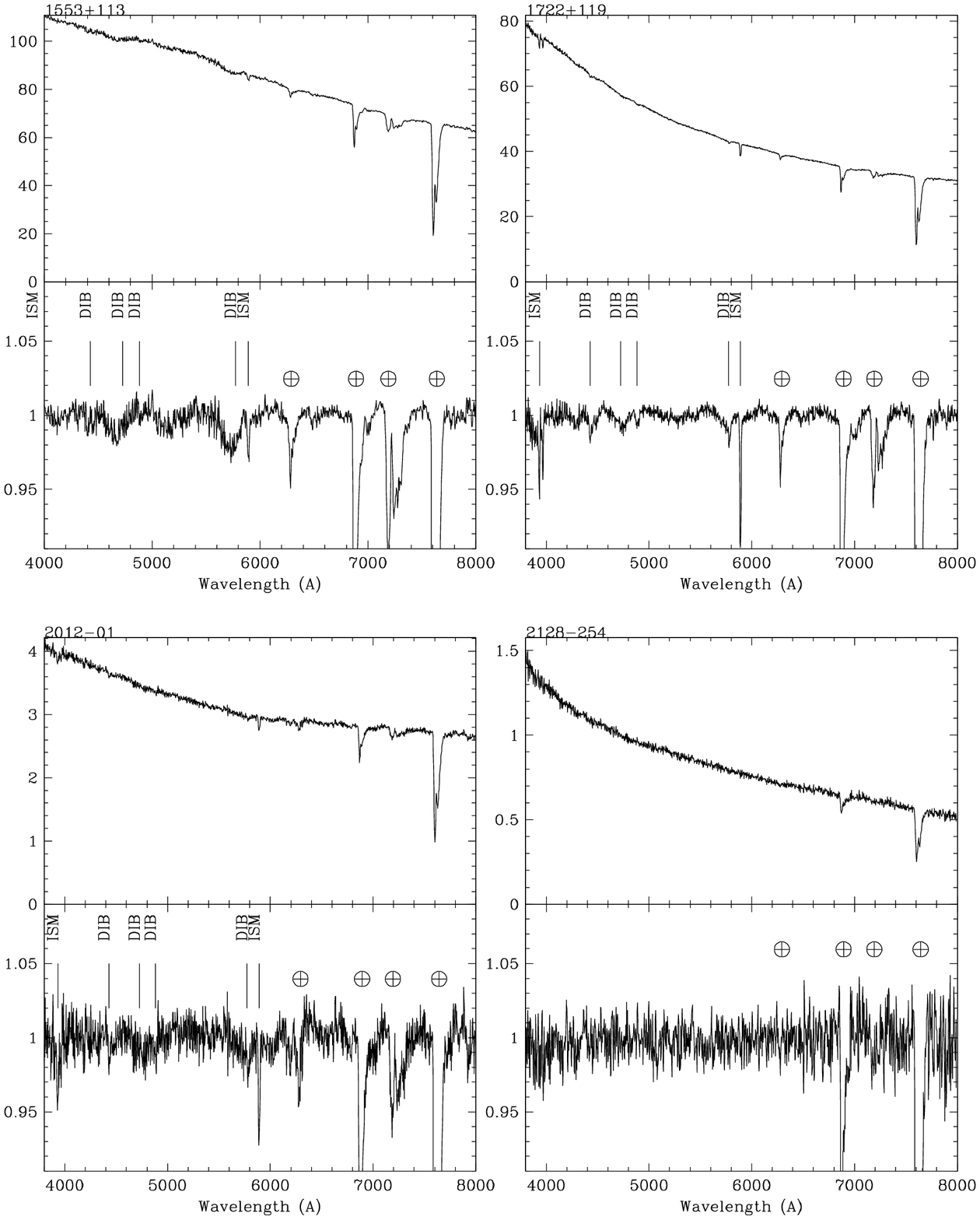}}
  \caption{continued}

\end{figure}
\addtocounter{figure}{-1}

\clearpage

\begin{figure}[htbp]
    \centering
  \resizebox{\hsize}{!}{\includegraphics{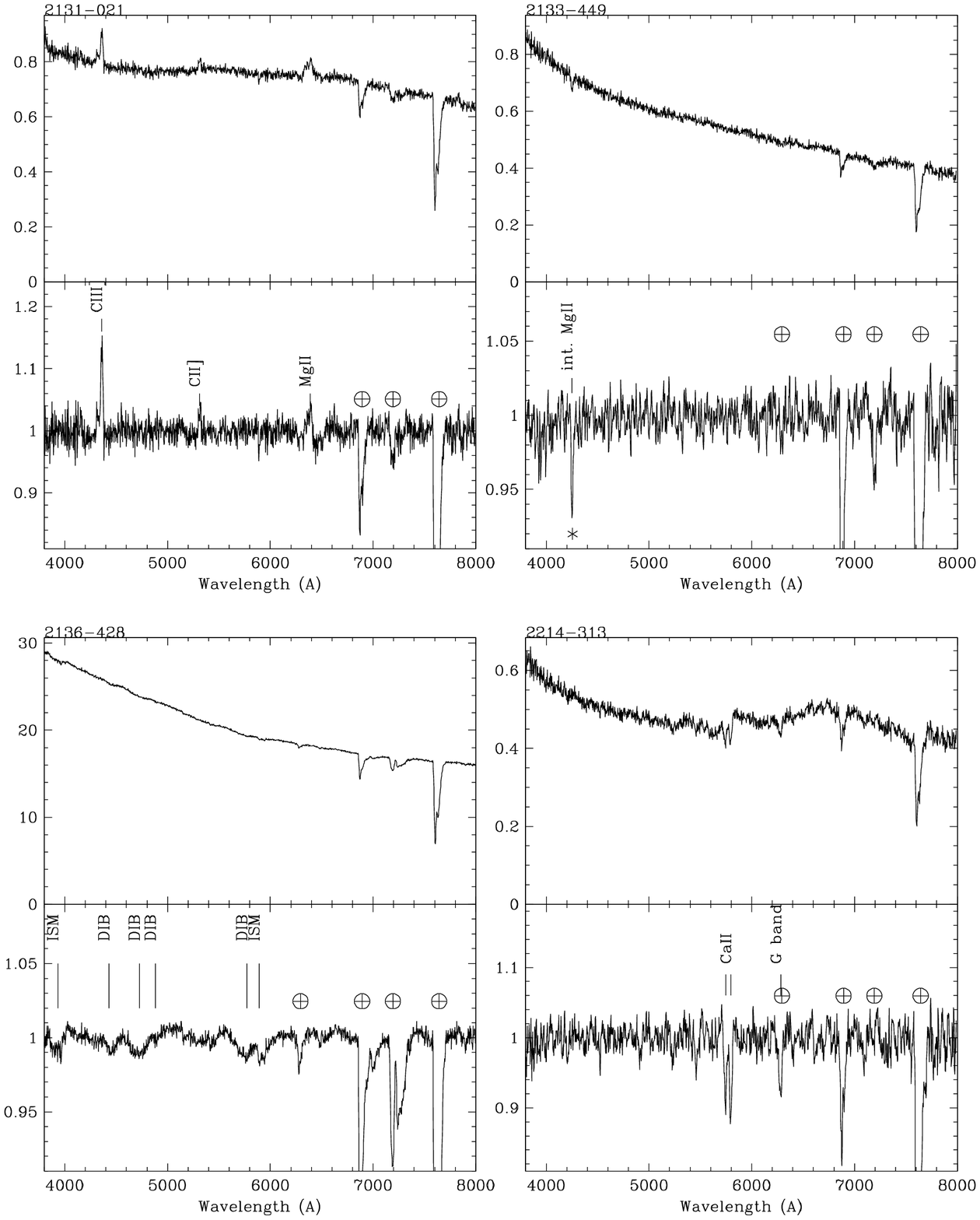}}
  \caption{continued}

\end{figure}
\addtocounter{figure}{-1}

\clearpage

\begin{figure}[htbp]
    \centering
  \resizebox{\hsize}{!}{\includegraphics{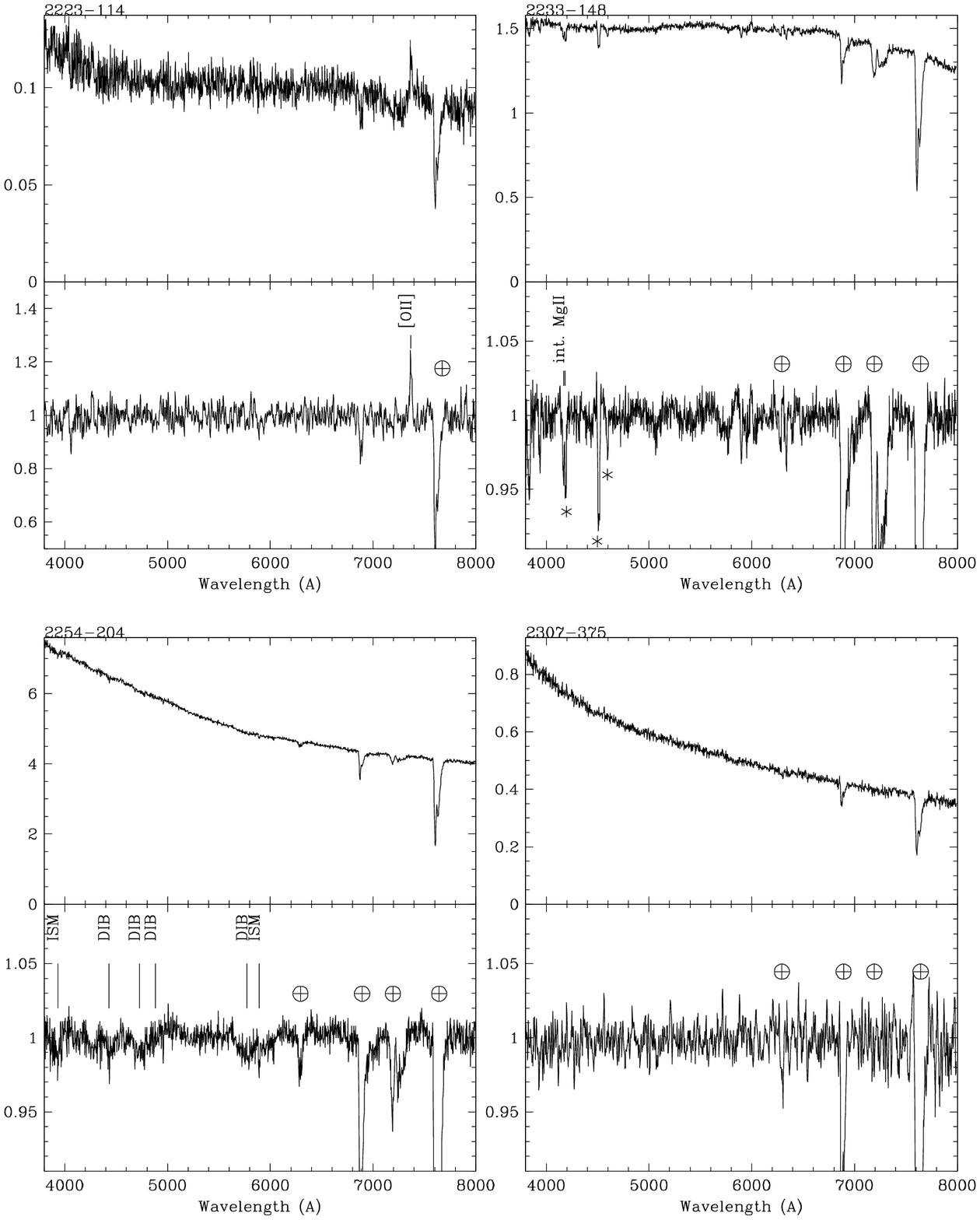}}
  \caption{continued}

\end{figure}
\addtocounter{figure}{-1}

\begin{figure}[htbp]
    \centering
  \resizebox{\hsize}{!}{\includegraphics{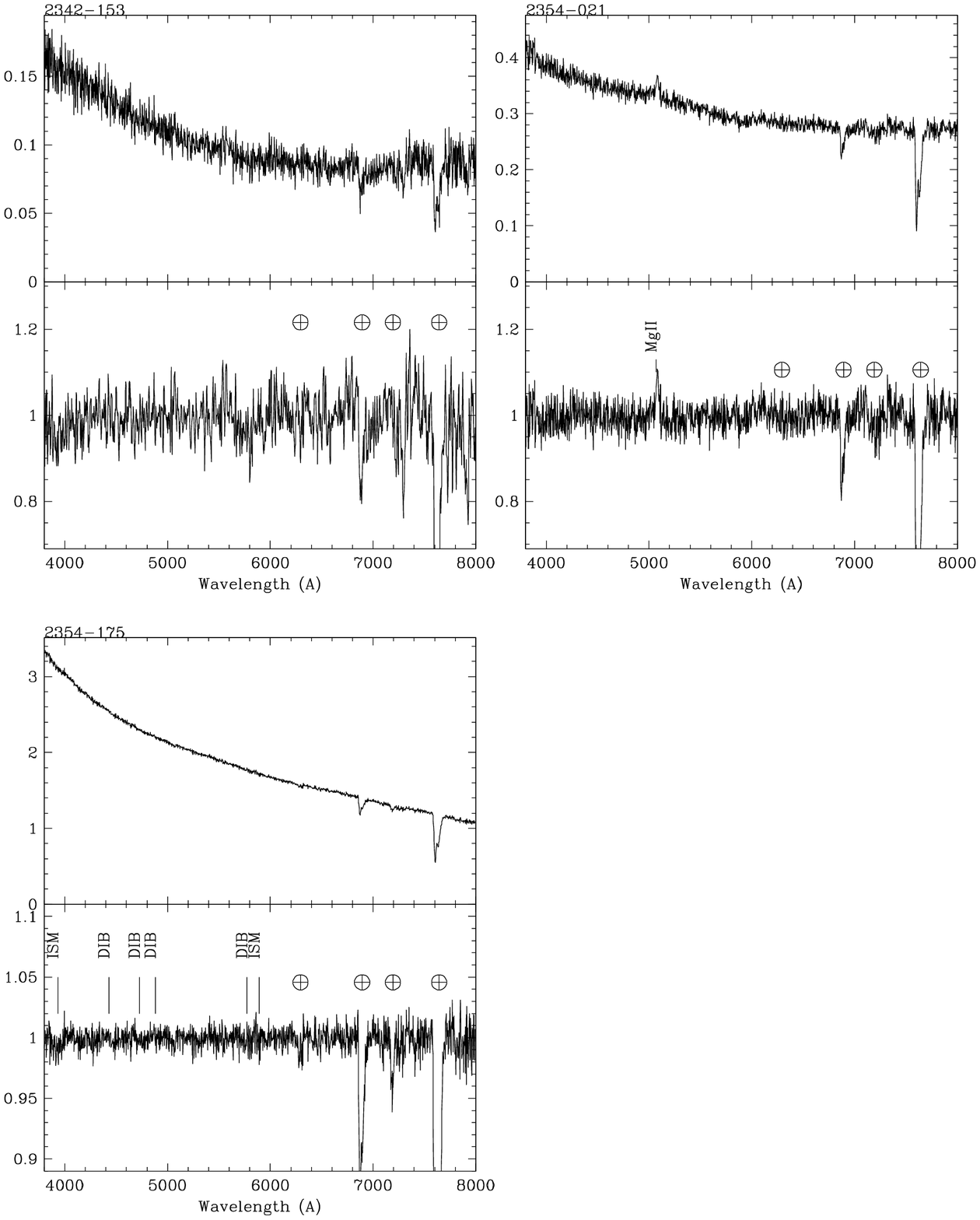}}
  \caption{continued}

\end{figure}

\clearpage

\begin{figure}[htbp]
    \centering
  \resizebox{\hsize}{!}{\includegraphics[scale=.9]{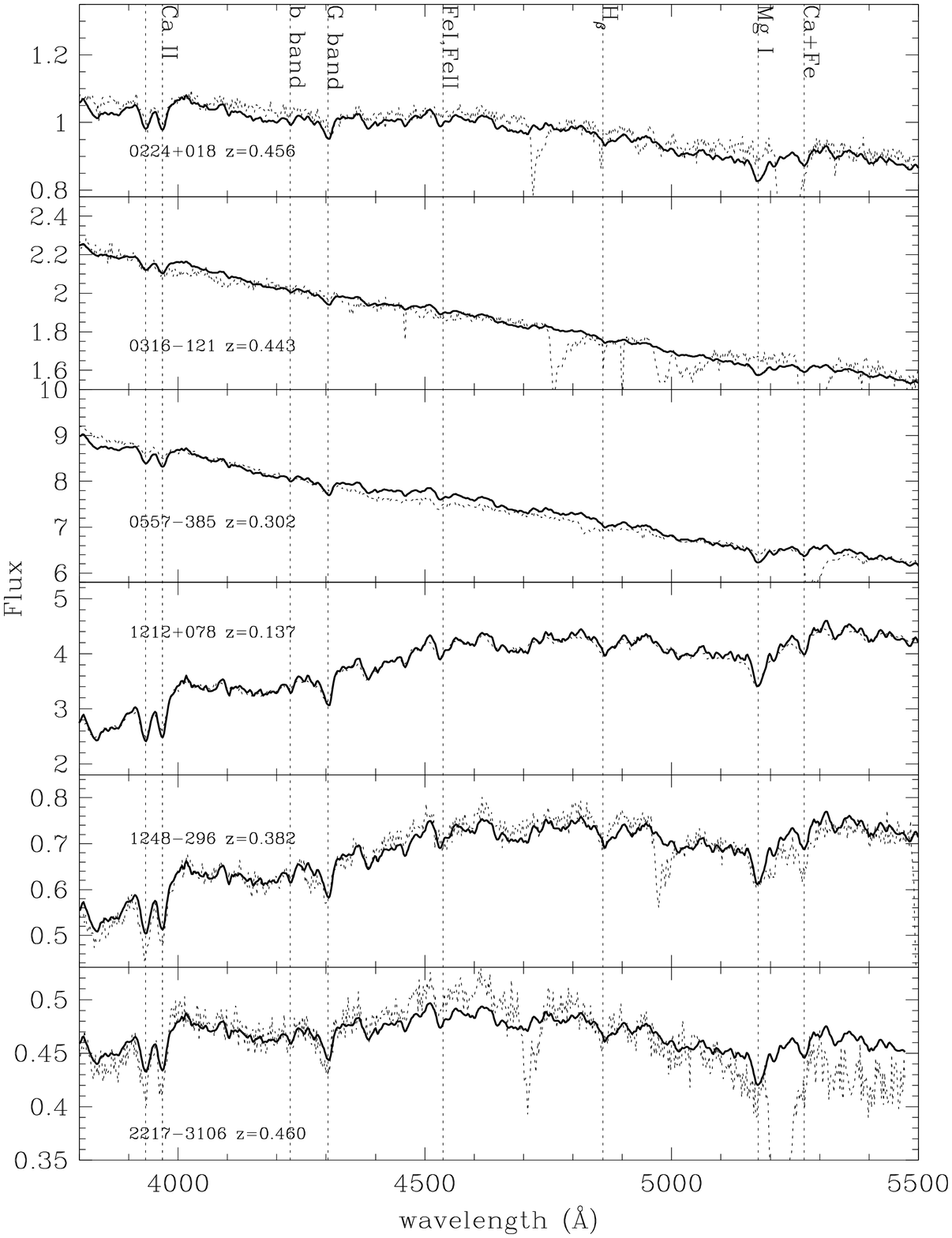}}
  \caption{Spectral decomposition for objects with detected host galaxy spectral
  features in the objects rest frame. Solid line shows the fitted spectrum, 
  dotted line the observed one. Objects 0224+018, 0316-121, 0557-385 were discussed in paper I.}
  \label{fig:gfit2}

\end{figure}

\begin{figure}[htbp]
    \centering
  \resizebox{\hsize}{!}{\includegraphics[scale=.9]{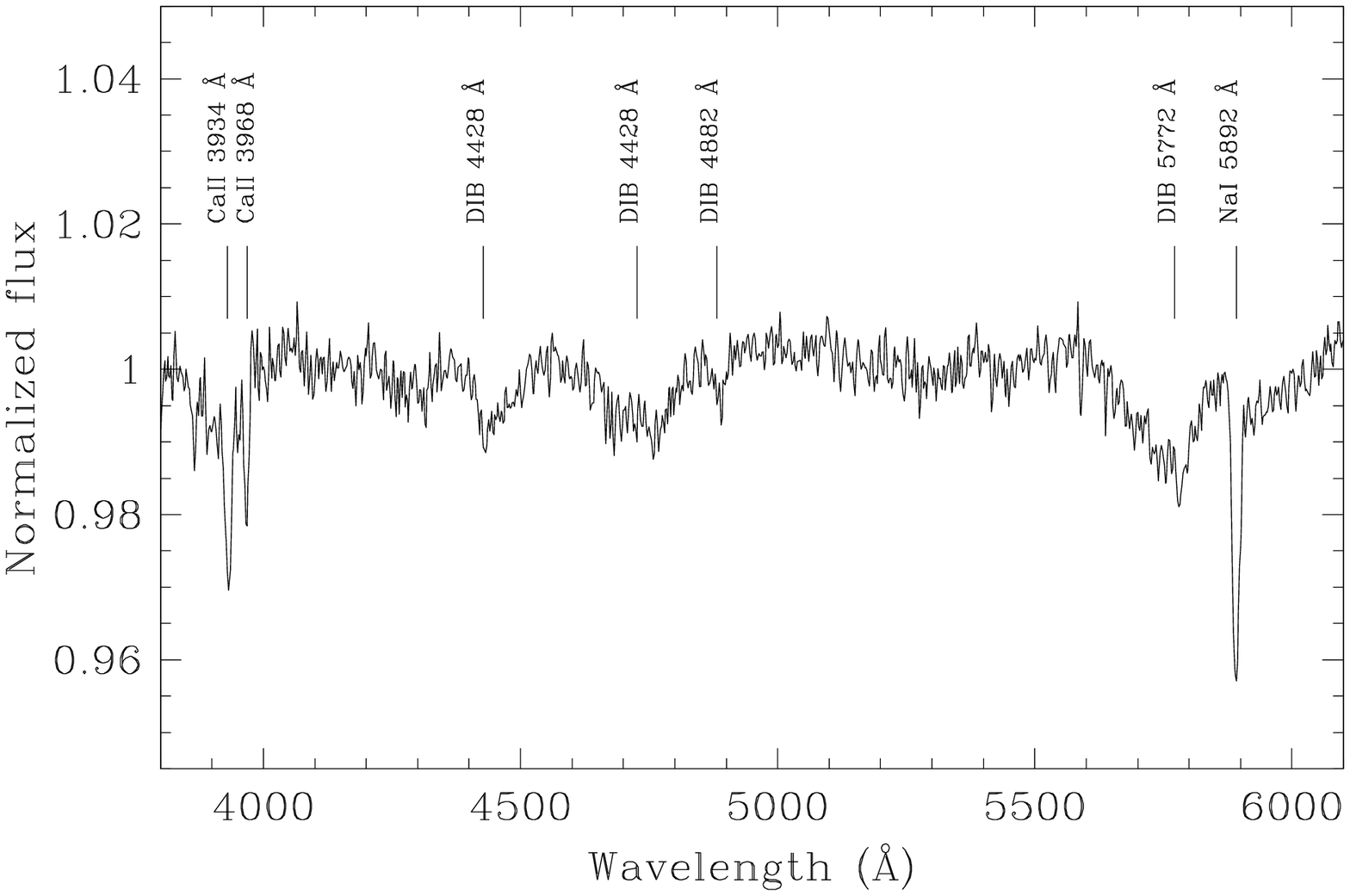}}
  \caption{Combined spectrum of the ISM features of 
  lineless BL Lacs.
  CaII $\lambda\lambda$3934,3968 and NaI $\lambda$5892 atomic lines
  and, diffuse interstellar bands (DIB) at
  $\lambda\lambda$4428,4726,4882,5772 are indicated}
  \label{fig:ISM}

\end{figure}
\clearpage

\begin{figure}[htbp]
    \centering
\includegraphics[scale=.7]{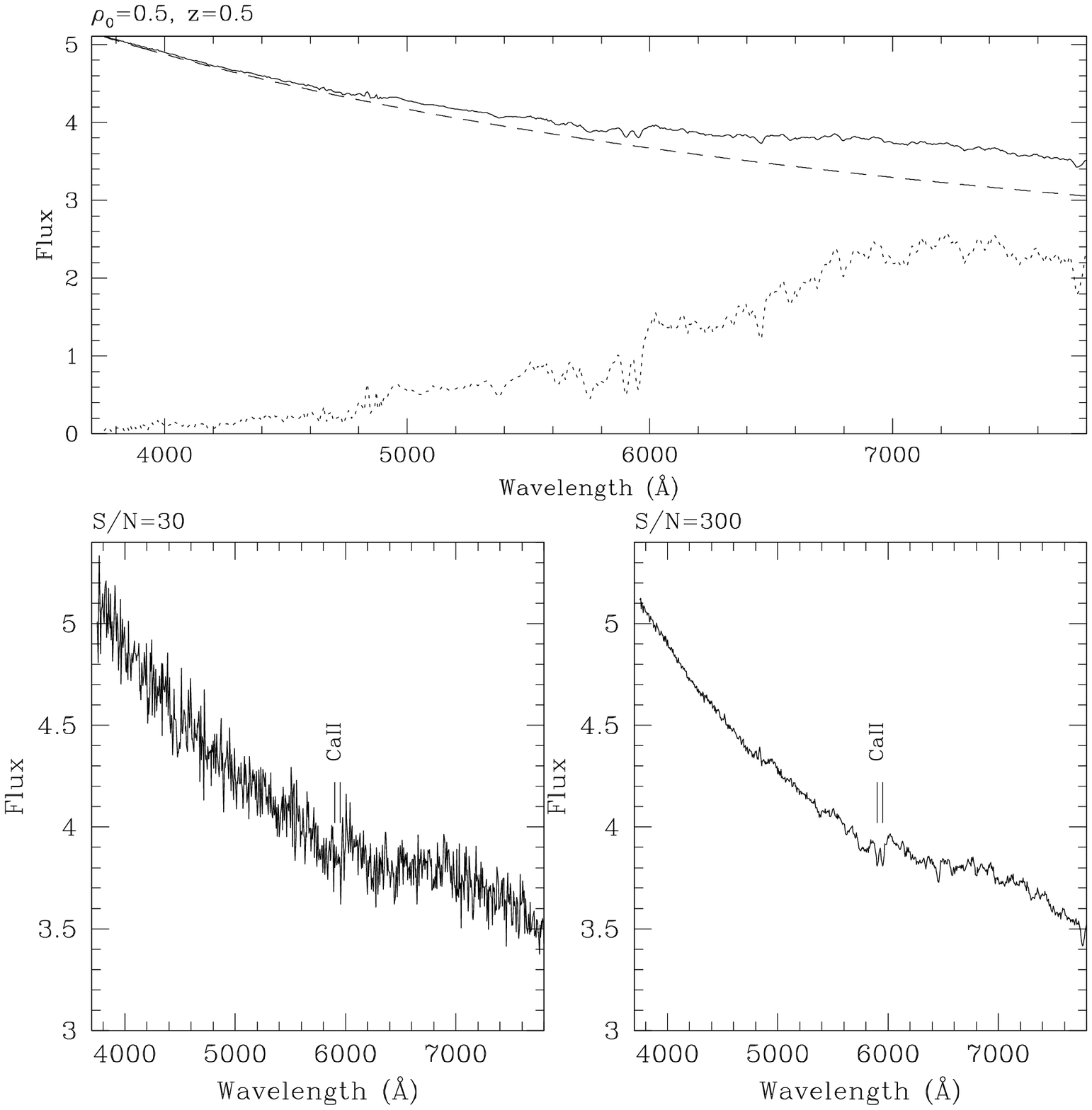}
  \caption{{\it Top panel:} Simulated BL Lac spectrum at z=0.5 
    (thick line), obtained as the composition of a non-thermal power-law 
    (dashed line) and an elliptical galaxy spectrum (magnified 5 times, dotted 
    line), with nucleus--to--host ratio of 5. {\it Bottom panels:} the 
    simulated spectrum, if observed with S/N=30 {\it (left)} or with S/N=300 
    {\it (right)}}
   \label{fig:specsim}
\end{figure}

\clearpage
\begin{figure}[htbp]
    \centering
\includegraphics[scale=.7]{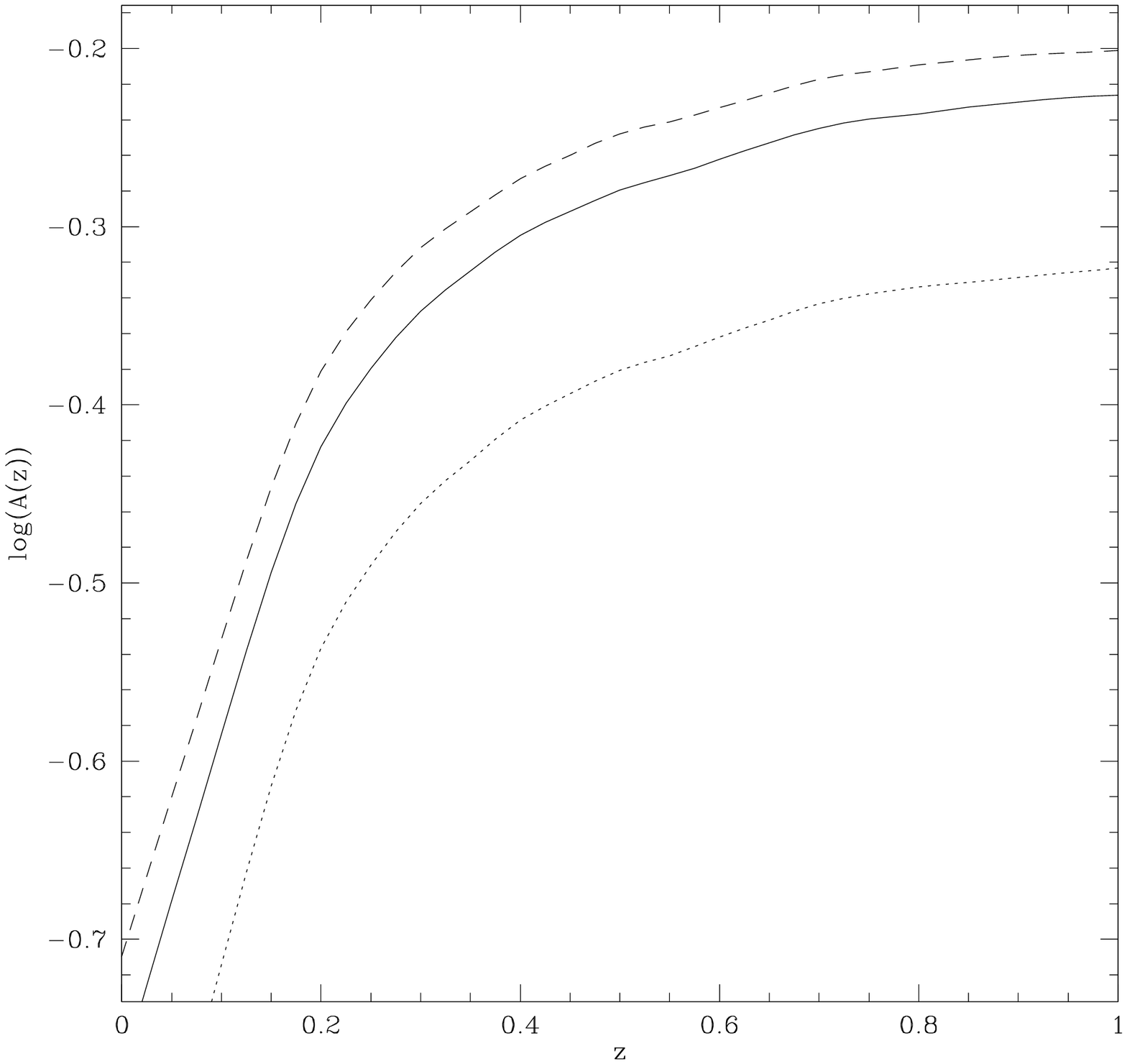}
  \caption{Aperture effect correction A(z) as a function of the
      redshift for aperture sizes 2''$\times$12'' (dashed), 2''$\times$6''
      (solid), 2''$\times$3'' (dotted). 
      Seeing is assumed to be $\sim$1''.} 
   \label{fig:apeffect}
\end{figure}

\clearpage

\begin{figure}[htbp]
    \centering
\includegraphics[scale=.7]{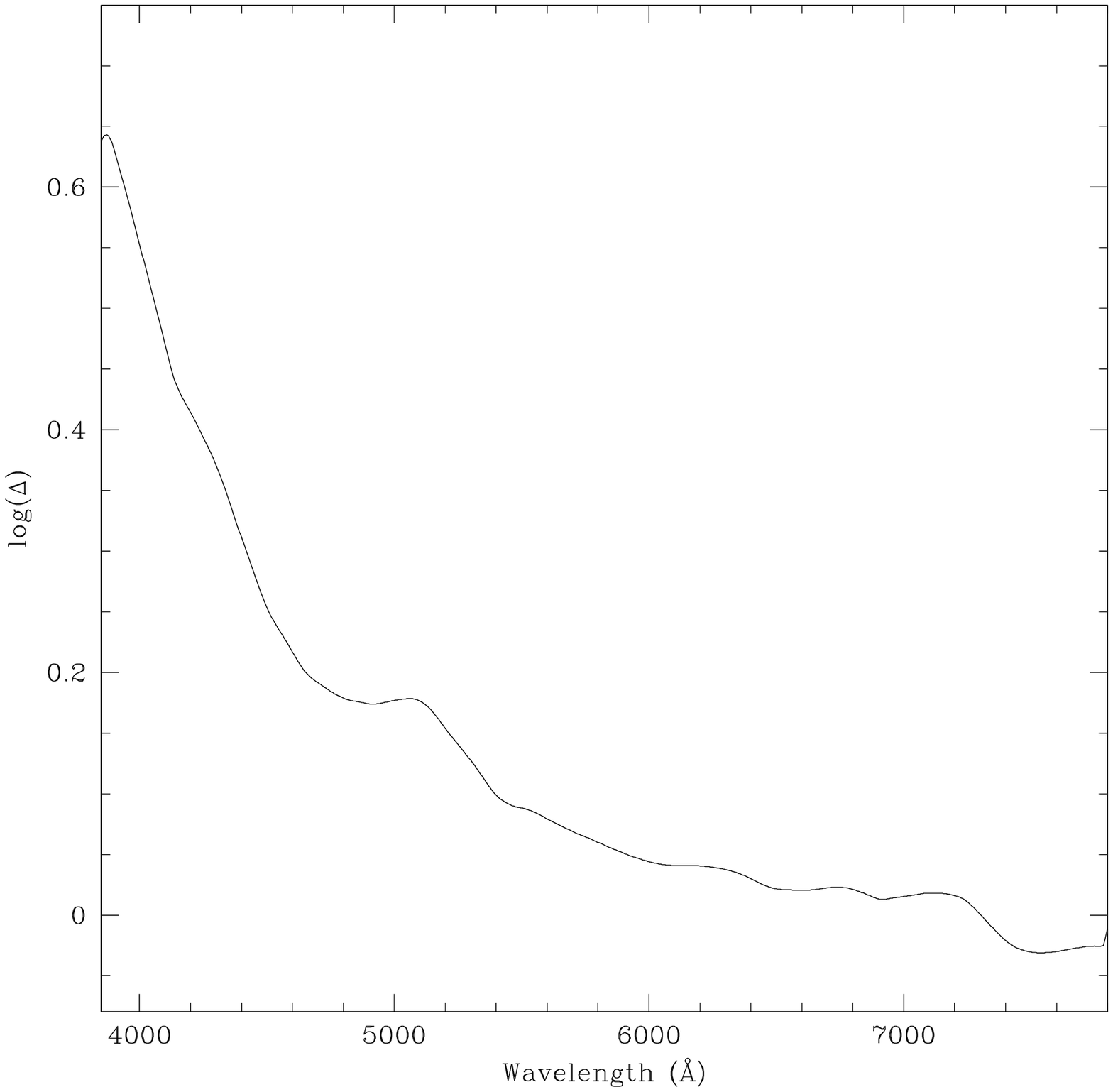}
  \caption{Relative nucleus--to--host ratio $\Delta$, referred to
      $\lambda_0$=6750 \AA (effective wavelength for R band magnitude), as a
      function of the wavelength. The assumed spectral index for the nuclear 
      component is $\alpha$=0.7. For CaII $\lambda$3934 absorption feature, 
      $\Delta$ equals 4.3.} 
   \label{fig:delta}
\end{figure}

\begin{figure}[htbp]
    \centering
\includegraphics[scale=.7]{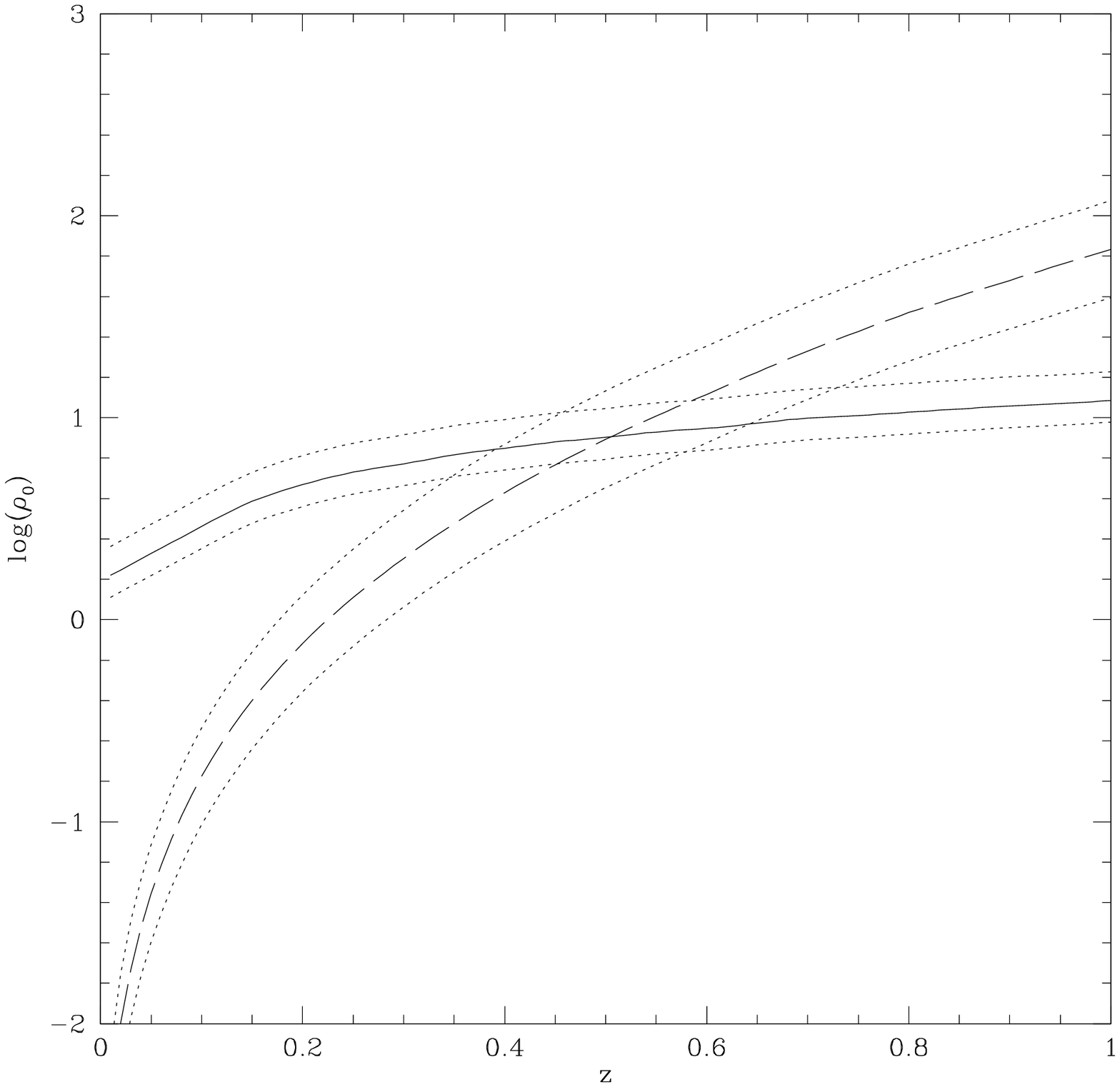}
  \caption{Redshift lower limit obtained from apparent magnitude and 
    \EWmin\ values for 1RXS J150343.0$-$154107. The thick solid line 
    represents the N/H vs z limit obtained from the \EWmin \ value. Dotted 
    curves correspond to a 0.1 \AA\ uncertainty on \EWmin. The dashed line 
    gives the N/H vs. z relation for a BL Lac with a host galaxy with 
    M$_R$=-22.9 and nuclear apparent magnitude R=17.7.  Dotted lines 
    correspond to the uncertainty due to the range of variation of host galaxy 
    magnitude (0.5 mag) and observational photometric errors (0.1 mag). 
    The intersection between the two solid lines gives 
    the lower limit on the redshift.  The analytic form of the 
    curves
    is described by Eqs. \ref{eq:ewnh}, \ref{eq:magnh}; further 
    details can
    be found in \citet{phd}}
   \label{fig:ew2z}
\end{figure}

\begin{figure}[htbp]
    \centering
  \resizebox{\hsize}{!}{\includegraphics[scale=.9]{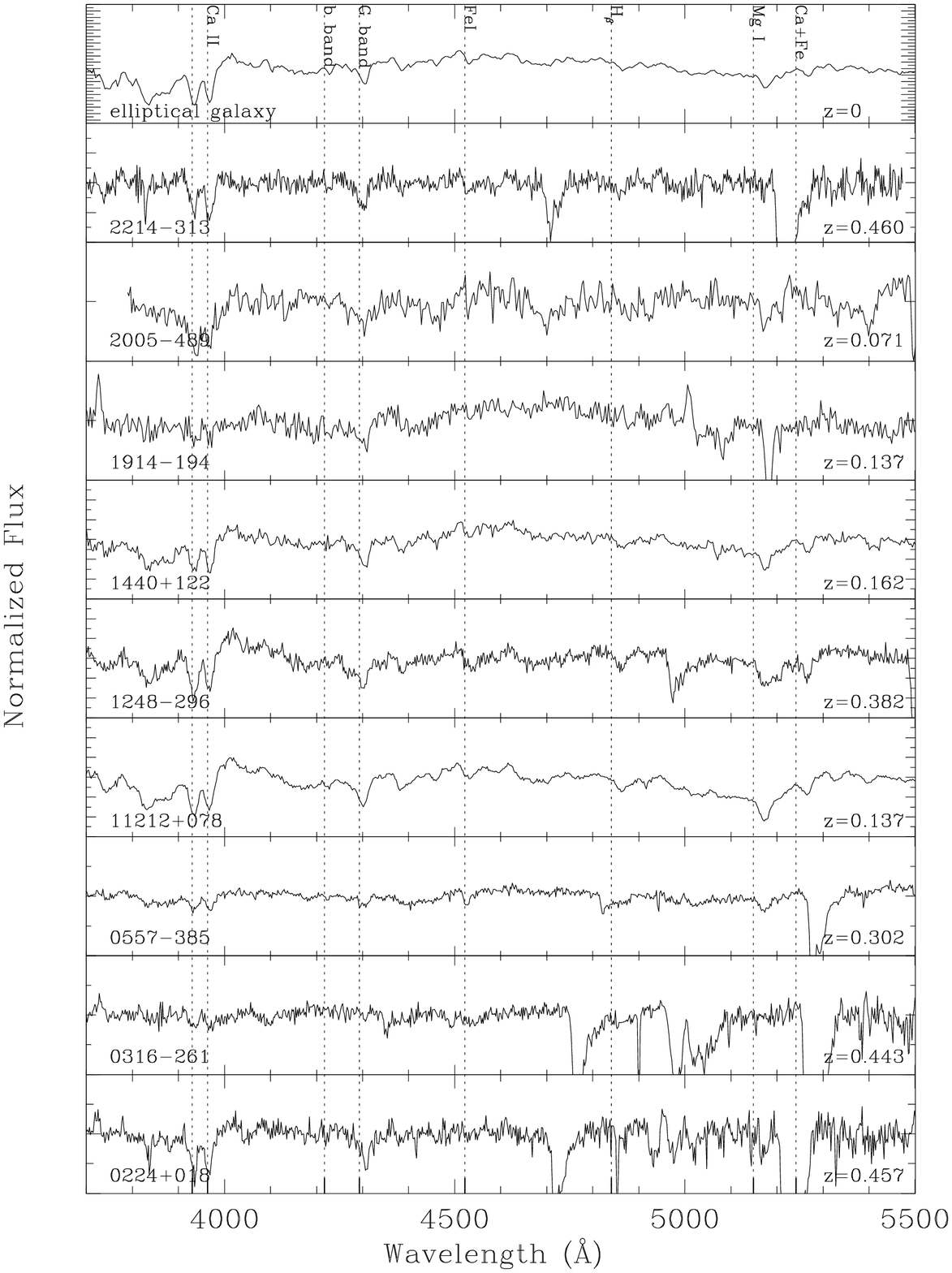}}
  \caption{ Rest Frame normalized spectra of objects used to compare redshift
    estimates from CaII equivalent widths with spectroscopic values.}
   \label{fig:zestspec}
\end{figure}

\clearpage

\begin{figure}[htbp]
    \centering
  \resizebox{\hsize}{!}{\includegraphics[scale=.9]{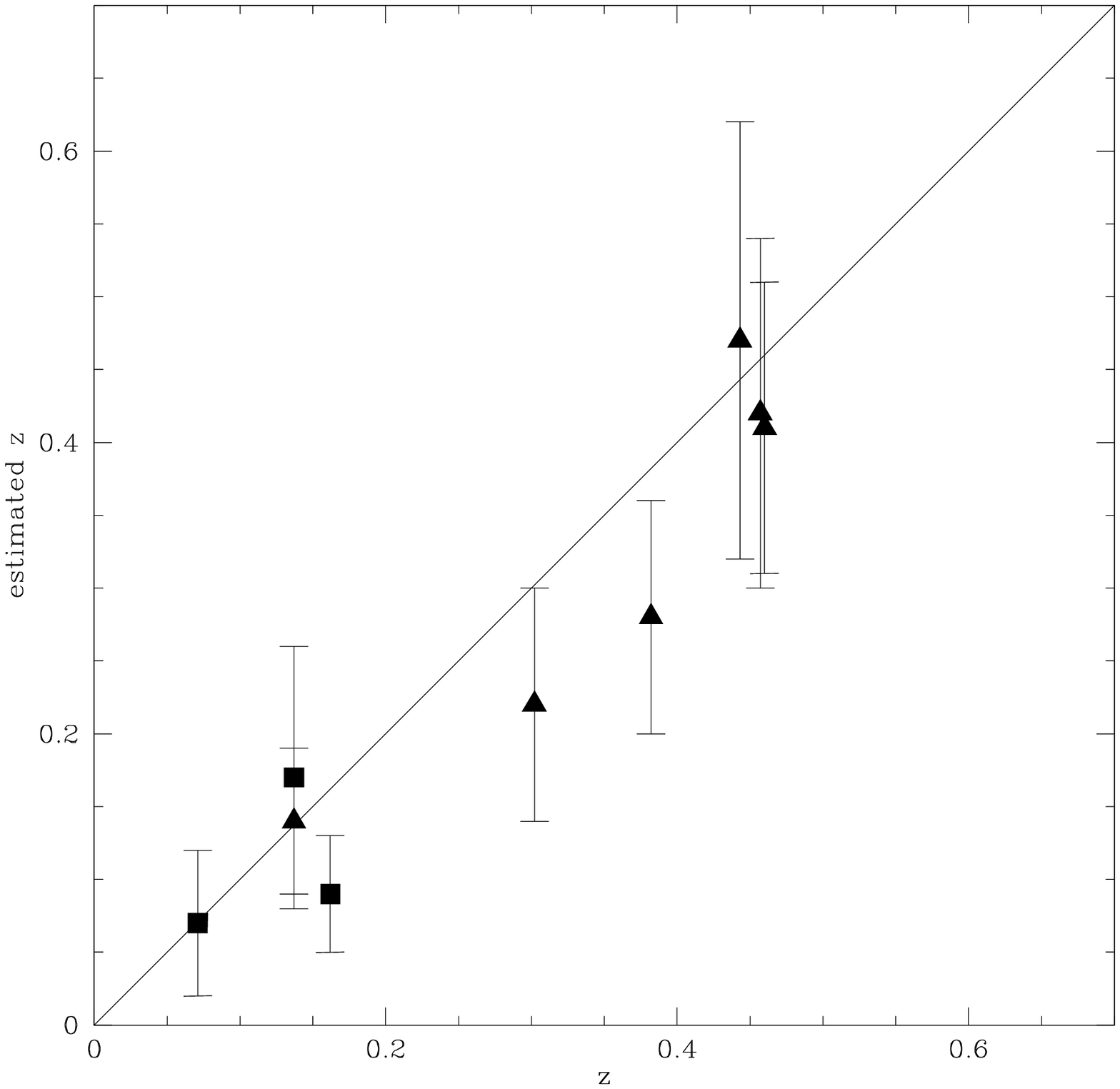}}
  \caption{Comparison between the spectroscopic redshift z and the value
    estimated from CaII equivalent widths (see section \ref{sec:ew2z} and 
  Fig. \ref{fig:ew2z}). Triangles 
  refer to objects observed at ESO 3.6m, squares to objects observed at 
  ESO VLT (this work).}
   \label{fig:zestcomp}
\end{figure}

\clearpage

\begin{figure}[htbp]
    \centering
  \resizebox{\hsize}{!}{\includegraphics[scale=.9]{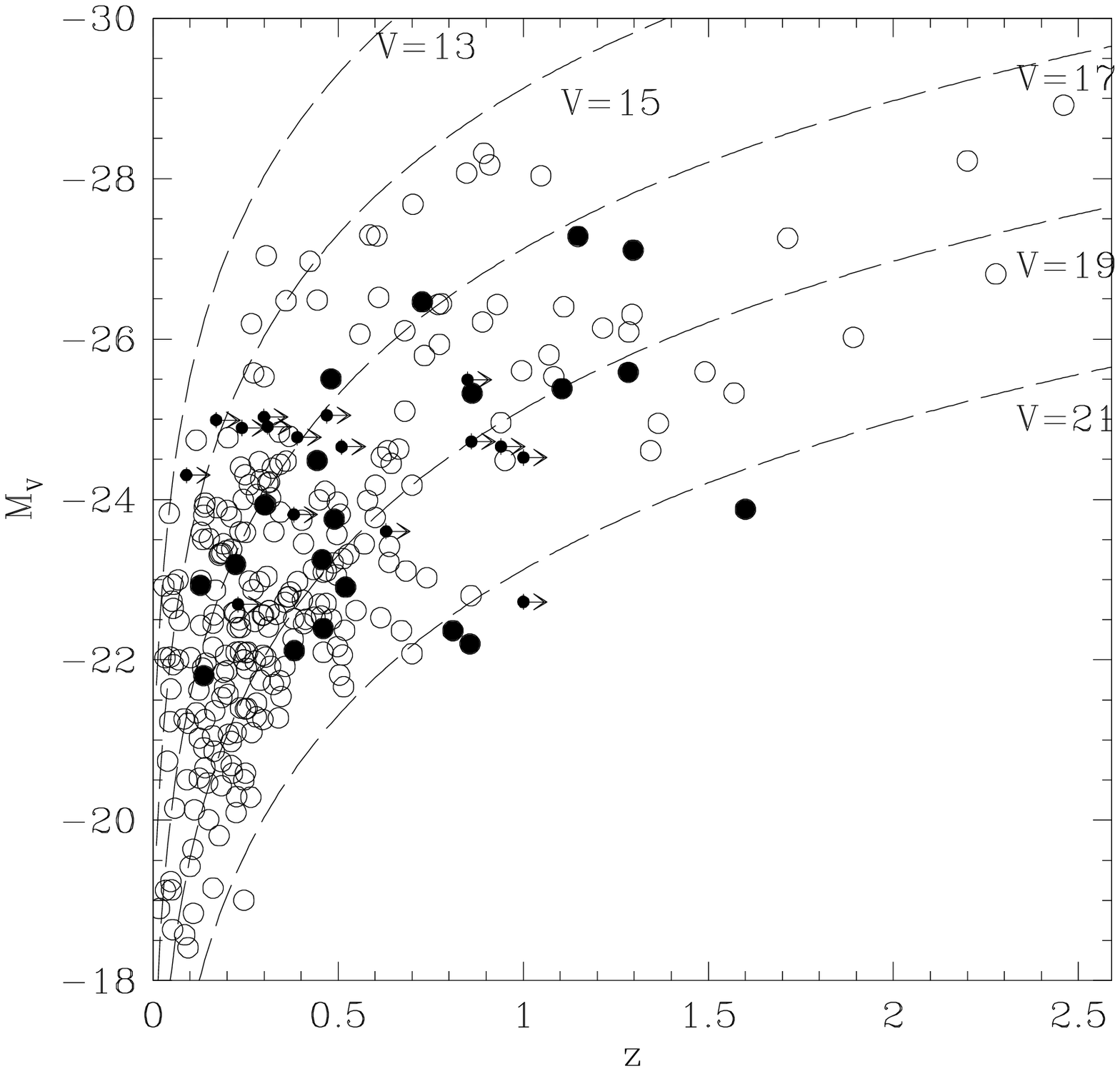}}
  \caption{M$_V$ vs. redshift distribution of : \citet{PG95} + SS sample with 
  redshift known from literature (open circles), BL Lacs the redshift of which 
  has been determined in paper I and in this work (filled circles), 
  including redshift lower limits (arrows).}
\label{fig:distz}
\end{figure}


\begin{deluxetable}{lllllllllcl}
\tabletypesize{\scriptsize}
\tablecaption{Journal of observations and results for objects not reported in 
paper I\label{tab:results}}
\tablehead{
\colhead{Object}&\colhead{IAU}&\colhead{RA}&\colhead{Dec}&\colhead{Date}&\colhead{t$_{exp}$}&\colhead{S/N}&
\colhead{R}&\colhead{$\alpha$}&\colhead{\EWmin}&
\colhead{z}\\
\colhead{name}&\colhead{name}&\colhead{(J2000)}&\colhead{(J2000)}&\colhead{}&\colhead{}&\colhead{}&
\colhead{}&\colhead{}&\colhead{}&
\colhead{}\\
\colhead{(1)}&\colhead{(2)}&\colhead{(3)}&\colhead{(4)}&\colhead{(5)}&\colhead{(6)}&\colhead{(7)}&
\colhead{(8)}&\colhead{(9)}&\colhead{(10)}&
\colhead{(11)}}
\startdata
PKS 0047+023		  & 0047+023	    & 00 49 43.2  &   +02 37 04.8  & 05 Aug 03& 1800 & 80  & 19.0 & 0.61 & 0.36& $>$0.82\\
PKS 0048$-$09		  & 0048$-$097      & 00 50 41.3  & $-$09 29 05.2  & 17 Sep 03& 1800 & 250 & 16.0 & 0.95 & 0.22& $>$0.30\\
PKS 0420+022		  & 0420+022	    & 04 22 52.2  &   +02 19 26.9  & 19 Nov 03& 2325 & 90  & 18.9 & *	 & 0.41&   2.278\\
PKS 0422+00		  & 0422+004	    & 04 24 46.8  &   +00 36 06.3  & 27 Nov 03& 2325 & 230 & 16.2 & 0.88 & 0.25& $>$0.31\\
PKS 0627$-$199  	  & 0627$-$199      & 06 29 23.8  & $-$19 59 19.7  & 16 Dec 03& 2325 & 50  & 19.3 & 0.56 & 0.92& $>$0.63\\
PKS 0723$-$00		  & 0723$-$008      & 07 25 50.6  & $-$00 54 56.5  & 25 Dec 03& 2325 & 250 & 16.0 & *	 & 0.23&   0.127\\
H 0841+1256		  & 0841+129	    & 08 44 24.1  &   +12 45 48.0  & 30 Dec 03& 2325 & 100 & 18.0 & *	 & 0.38& $>$2.48\\
HB89 1210+121		  & 1210+121	    & 12 12 33.9  &   +11 50 56.9  & 24 Jan 04& 2325 & 180 & 17.8 & *	 & 0.28&   *	\\
1ES 1212+078		  & 1212+078	    & 12 15 10.9  &   +07 32 03.8  & 25 Jan 04& 2325 & 100 & 17.3 & 1.17 & 0.39&   0.137\\
1222+102		  & 1222+102	    & 12 25 23.1  &   +09 59 35.0  & 26 Jan 04& 2325 & 160 & 17.7 & 2.67 & 0.30&   *	\\
1ES 1248$-$296		  & 1248$-$296      & 12 51 34.9  & $-$29 58 42.9  & 24 Jan 04& 2325 & 50  & 19.5 & 0.92 & 0.57&   0.382\\
UM566			  & 1319+019	    & 13 19 55.1  &   +01 52 58.3  & 30 Apr 03& 1800 & 100 & 18.2 & *	 & 0.36&   *	\\
1ES 1320+084N		  & 1320+084	    & 13 22 54.9  &   +08 10 10.0  & 30 Apr 03& 2325 & 50  & 19.5 & *	 & 0.54&   1.500\\
PKS 1349$-$439  	  & 1349$-$439      & 13 52 56.5  & $-$44 12 40.4  & 30 Apr 03& 1800 & 240 & 16.9 & 0.82 & 0.32& $>$0.39\\
1RXS J144505.9$-$032613   & 1442$-$032      & 14 45 05.8  & $-$03 26 12.8  & 28 Aug 04& 2325 & 100 & 17.7 & 1.21 & 0.35& $>$0.51\\
1RXS J150343.0$-$154107   & 1500$-$154      & 15 03 42.9  & $-$15 41 07.0  & 28 Aug 04& 2325 & 40  & 17.8 & 1.52 & 0.78& $>$0.38\\
HB89 1553+113		  & 1553+113	    & 15 55 43.0  &   +11 11 24.4  & 01 Aug 03& 1800 & 250 & 14.0 & 0.84 & 0.25& $>$0.09\\
H 1722+119		  & 1722+119	    & 17 25 04.4  &   +11 52 15.2  & 06 Apr 03& 1800 & 350 & 14.7 & 1.30 & 0.18& $>$0.17\\
PKS 2012$-$017  	  & 2012$-$017      & 20 15 15.2  & $-$01 37 33.0  & 31 Jul 03& 1800 & 130 & 19.3 & 0.49 & 0.34& $>$0.94\\
1RXS J213151.7$-$251602   & 2128$-$254      & 21 31 51.6  & $-$25 16 00.8  & 10 Jul 04& 2325 & 70  & 19.0 & 1.28 & 0.32& $>$0.86\\
PKS 2131$-$021  	  & 2131$-$021      & 21 34 10.3  & $-$01 53 17.0  & 18 Jul 04& 2325 & 80  & 19.2 & 0.29 & 0.43&   1.284\\
MH 2133$-$449		  & 2133$-$449      & 21 36 18.4  & $-$44 43 49.0  & 12 Jul 04& 2325 & 60  & 19.5 & 1.02 & 0.37& $>$0.98\\
MH 2136$-$428		  & 2136$-$428      & 21 39 24.1  & $-$42 35 21.3  & 03 Jul 03& 1800 & 490 & 15.6 & 0.84 & 0.24& $>$0.24\\
RX J22174$-$3106  	  & 2214$-$313      & 22 17 28.4  & $-$31 06 19.0  & 10 Jul 04& 2325 & 50  & 19.7 & 0.90 & 0.68&   0.460\\
PKS 2223$-$114  	  & 2223$-$114      & 22 25 43.6  & $-$11 13 40.0  & 02 Sep 04& 2325 & 20  & 21.5 & 0.31 & 1.04&   0.997\\
PKS 2233$-$148  	  & 2233$-$148      & 22 36 34.0  & $-$14 33 21.0  & 02 Sep 04& 2325 & 170 & 18.5 & 0.15 & 0.30& $>$0.65\\
PKS 2254$-$204  	  & 2254$-$204      & 22 56 41.2  & $-$20 11 40.3  & 31 Jul 03& 1800 & 220 & 17.1 & 0.86 & 0.25& $>$0.47\\
1RXS J231027.0$-$371926	  & 2307$-$375      & 23 10 26.9  & $-$37 19 26.0  & 10 Jul 04& 2325 & 80  & 19.6 & 1.15 & 0.34& $>$1.03\\
MS 2342.7$-$1531  	  & 2342$-$153      & 23 45 22.4  & $-$15 15 06.7  & 26 Jul 03& 2325 & 20  & 21.4 & 1.02 & 1.72& $>$1.03\\
1RXS J235730.1$-$171801	  & 2354$-$175      & 23 57 29.7  & $-$17 18 05.3  & 12 Jul 04& 2325 & 150 & 18.2 & 1.44 & 0.17& $>$0.85\\
\enddata	     
\tablecomments{Description of columns: 
(1) Object name; 
(2) IAU name;
(3) Right Ascension (J2000); 
(4) Declination (J2000); 
(5) Date of observations; 
(6) Exposure time (seconds); 
(7) Signal to Noise;
(8) Seeing during observations;
(9) Spectral index of the continuum, $\alpha$, defined by 
F$_{\lambda}\propto\lambda^{-\alpha}$;
(10) Minimum detectable EW;
(11) Redshifts measured from spectral features and redshift lower 
     limits from the procedure described in section \ref{sec:resb}.}
\end{deluxetable}

\begin{deluxetable}{llllllll}
\tabletypesize{\scriptsize}
\tablecaption{Parameters of BL Lac spectral decomposition\label{tab:gfit}}
\tablehead{
\colhead{Object}&\colhead{z}&\colhead{$\alpha$}&\colhead{m$_R^{host}$}
&\colhead{M$_R^{host}$}&\colhead{$\rho_0$}&\colhead{M$^{host}_{phot}$}&\colhead{Note}\\
\colhead{(1)}&\colhead{(2)}&\colhead{(3)}&\colhead{(4)}&\colhead{(5)}&
\colhead{(6)}&\colhead{(7)}&\colhead{(8)}\\}
\startdata
0224+018	&0.456&	1.50&	19.9&	$-$23.2	&2.2	&$-$23.1	&(1)\\
0316$-$121	&0.443&	1.44&	20.2&	$-$22.8	&6.4	&		&(1)\\
0557$-$385	&0.302&	1.61&	18.3&	$-$23.4	&5.5	&		&(1)\\
1212+078	&0.137& 1.17&   17.4&   $-$22.0	&0.4	&$-$23.0	&   \\
1248$-$296	&0.382& 0.92&   19.7&   $-$22.7	&0.8	&$-$23.7	&   \\
2214$-$313	&0.460&	0.90&	20.8&	$-$22.3	&2.1	&		&   \\
\enddata             
\tablecomments{Description of columns:
(1) Object name
(2) Redshift
(3) Fitted spectral index of the continuum, $\alpha$, defined by 
F$_{\lambda}\propto\lambda^{-\alpha}$;
(4) fitted R magnitude of the host galaxy;
(5) Absolute R magnitude of the host galaxy, corrected for extinction and
evolution, but not for aperture effects;
(6) Rest frame R band nucleus-to-host flux ratio;
(7) Absolute R magnitude of the host galaxy from photometry;
(8) Note: (1) Spectrum published in \citet{sbarufatti05a}.}
\end{deluxetable}

\begin{deluxetable}{llllllllr}
\tabletypesize{\scriptsize}
\tablecaption{Measurements of spectral lines\label{tab:bllines}}
\tablehead{
\colhead{Object name}&\colhead{Object class}&\colhead{z$_{avg}$}&\colhead{Line ID}&\colhead{$\lambda$}&\colhead{z}&\colhead{Type}&\colhead{FWHM}&\colhead{EW}\\
\colhead{(1)}&\colhead{(2)}&\colhead{(3)}&\colhead{(4)}&\colhead{(5)}&\colhead{(6)}&\colhead{(7)}&\colhead{(8)}&\colhead{(9)}\\}
\startdata
0420+022	    &  QSO   &  2.278	&       &	&       &	&	&	\\
		    &	     &	      	&Ly$_{alpha}$& 4020 & 2.278& e   	&9400   &-73.0    \\
		    &	     &	      	&SiII   & 4285  &  2.279&  e   	&3900   &-4.0     \\
		    &	     &	      	&CII    & 4381  &  2.281&  e   	&5500   &-3.0     \\
		    &	     &	      	&SiIV   & 4587  &  2.283&  e   	&6300   &-24.0    \\
		    &	     &	      	&CIV    & 5077  &  2.278&  e   	&4900   &-50.0    \\
		    &	     &	      	&CIII]  & 6250  &  2.274&  e   	&4500   &-45.0    \\      
0723$-$008	    & QSO/BLL&  0.127  	&	&	&	&   	&	&	\\
                    &	     &  	&[NeV]  &  3858 & 0.126 &  e	&  1100 & -0.4  \\
		    &	     &	 	&[OII]  &  4200 & 0.127 &  e	&  1200 & -2.1  \\
		    &	     &	 	&[NeIII]&  4359 & 0.127 &  e	&  1300 & -0.7  \\
                    &        & 		&CaII   &  4433 & 0.127	&  g	&	&  0.6	\\
                    &        & 		&b band &  4477 & 0.127	&  g	&	&  0.1	\\
                    &        & 		&G band &  4847 & 0.127	&  g	&	&  0.1	\\
		    &	     &	 	&H$_{\gamma}$ & 4897 & 0.128 & e&  3700 & -1.0  \\
		    &	     &	 	&H$_{\beta}$ &  5477 & 0.127 & e&  1100 & -1.0  \\
		    &	     &	 	&[OIII] &  5587 & 0.127 &  e	&  900  & -1.9  \\
		    &	     &	 	&[OIII] &  5642 & 0.127 &  e	&  900  & -5.5  \\
                    &        & 		&Mg I   &  5830 & 0.127	&  g	&	&  0.4	\\
                    &        & 		&NaI    &  6645 & 0.127	&  g	&	&  0.1	\\
		    &	     &	 	&[OII]  &  7098 & 0.127 &  e	&  1000 & -1.2  \\
		    &	     &	 	&H$_{\alpha}$&  7402 & 0.128 & e&  1900 & -8.7  \\
		    &	     &	 	&SII	&  7561 & 0.126 &  e	&  600  & -1.1  \\
		    &	     &	 	&	&	&	&	&	&	\\
1212+078            &  BLL   &  0.137  	&       &     	&	&	&       &       \\
                    &        &         	& CaII  & 4473	& 0.137 &  g	&       &  5.7 	\\
                    &        &         	& CaII  & 4510	& 0.137 &  g	&       &  4.8 	\\
                    &        &         	& G band& 4890	& 0.136 &  g	&       &  5.3 	\\
                    &        &         	&H$_{\beta}$&5529&0.137 &  g	&       &  3.5 	\\
                    &        &         	& Mg I  & 5883	&0.137 	&  g	&	& 16.2 	\\
                    &        &         	& Na I  & 6696	&0.137 	&  g	&	&  3.9 	\\
                    &        &         	&H$_{\alpha}$&7481&0.139&  e	&  700  & -2.0 	\\
1248$-$296          &  BLL   &   0.382 	&       &       &       &   	&	&      	\\
                    &        &         	& CaII  &  5436	& 0.382	&  g	& 2100  &  6.1 	\\
                    &        &         	& CaII  &  5482	&	&  g	& 1700  &  4.2 	\\
                    &        &         	& G band&  5942	&	&  g	& 2700  &  5.6 	\\
                    &        &         	&H$_{\beta}$&6719&0.382 &  g	& 1600  &  1.5 	\\
\tablebreak
1320+084            &  QSO   &   1.500 	&       &  	&	&      	&	&      	\\
		    &	     &		& CIV	&  3873	& 1.500	&  e	& 2200	&  1.9	\\
		    &	     &		& HeII	&  4095	& 1.500	&  e	& 1900	& -9.0	\\
		    &        & 		& ?     &  4234	& 	&  a	&	&-80.2	\\
		    &	     &		& NIII]	&  4376	& 1.500	&  e	& 2300	& -5.8	\\
		    &	     &		& CIII]	&  4770	& 1.500	&  e	& 2200	&-40.9	\\
		    &        & 		& ?     &  6071	& 	&  a	&	&  0.6	\\
		    &        & 		& ?     &  6130	& 	&  a	&	&  1.4	\\
		    &        & 		& MgII  &  6563 & 1.347	&  a	&	&  1.8	\\
		    &        & 		& MgII  &  6578 & 1.347	&  a	&	&  1.6	\\
		    &	     &		& MgII	&  7001	& 1.502	&  e	& 6700	&-60.0	\\
2131$-$021          &  BLL   &  1.283  	&       &  	&	&      	&	&      	\\
                    &        &         	&  CIII]&  4357 & 1.283 &  e   	& 2000  & -4.4 	\\
                    &        &         	&  CII] &  5312 & 1.284 &  e   	& 700   & -1.4 	\\
                    &        &         	&  MgII &  6383 & 1.281 &  e   	& 3000  & -3.8 	\\
2133$-$449          &  BLL   &$>$0.52 	&       &  	&	&      	&       &    	\\
                    &        &         	&  MgII &  4250 & 0.519 &  a   	& 2500	& 1.5	\\
                    &        &         	&       &  	&	&      	&	&    	\\
2214$-$313          &  BLL   &  0.460  	&       &  	&	&      	&	&    	\\
                    &        &         	&  CaII &  5746 & 0.461 &  g   	& 3000  &  3.7 	\\
                    &        &         	&  CaII &  5792 & 0.460 &  g   	& 4300  &  3.3 	\\
                    &        &         	&  Gband&  6280 & 0.459 &  g   	& 3200  &  2.6 	\\
                    &        &         	&       &  	&	&      	&	&      	\\
2223$-$114          &  BLL   &  0.977   &       &  	&	&      	&	&      	\\
                    &        &         	& [OII] &  7367 &	&  e   	& 1200	& -5.0  \\
                    &        &         	&       &  	&	&      	&	&      	\\
2233$-$148          &  BLL   &$>$0.49   &       &  	&	&      	&	&      	\\
                    &        &         	&  MgII &  4165 & 0.490 &  a   	&	& 0.7  	\\
                    &        &         	&  MgII &  4183 & 0.493 &  a   	&	& 0.9  	\\
                    &        &         	&  ?    &  4514 &	&  a   	&	& 1.7  	\\
                    &        &         	&  ?    &  4598 &	&  a   	&	& 0.5  	\\
                    &        &         	&       &  	&	&      	&	&      	\\
\enddata	
\tablecomments{Description of columns: 
(1) Object name;     
(2) Object class; 
(3) average redshift;
(4) line identification; 
(5) observed wavelength of line center(\AA); 
(6) redshift of the line; 
(7) type of the line (\textbf{e}: emission line; 
\textbf{g}: absorption line from the host galaxy;
\textbf{a}: absorption line from intervening systems); 
(8) FWHM of the line (km s$^{-1}$); 
(9) EW of the line (\AA); 
}
\end{deluxetable}

\begin{deluxetable}{lllllll}
\tabletypesize{\scriptsize}
\tablecaption{Comparison between spectroscopic redshifts and estimates
  obtained from \EWmin\ values.\label{tab:resultscomp}}
\tablehead{
\colhead{Object name}&
\colhead{z}&
\colhead{m$_{R}^{nucleus}$}&
\colhead{EW$_{CaII}$}&
\colhead{z$_{est}$}&
\colhead{$\delta_{z_{est}}$}&
\colhead{Tel.}\\
\colhead{(1)}&
\colhead{(2)}&
\colhead{(3)}&
\colhead{(4)}&
\colhead{(5)}&
\colhead{(6)}&
\colhead{(7)}\\
}
\startdata
0224+018   	&0.457   &19.0*   &1.7   &0.42    &0.12  &VLT	\\
0316$-$261    	&0.443   &18.1*   &0.6   &0.47    &0.15  &VLT	\\
0557$-$385   	&0.302   &16.9*   &0.9   &0.22    &0.08  &VLT	\\
1212+078	&0.137	 &18.5*   &5.7	 &0.14	  &0.05  &VLT	\\
1248$-$296	&0.382	 &19.8*   &6.1	 &0.28	  &0.08  &VLT	\\
1440+122	&0.162   &17.2    &3.5   &0.09    &0.04  &ESO3.6\\
1914$-$194	&0.137   &15.8    &0.5   &0.17    &0.09  &ESO3.6\\
2005$-$489	&0.071   &14.1    &0.4   &0.07    &0.05  &ESO3.6\\
2214$-$313	&0.460   &19.9*   &3.7   &0.41    &0.10  &VLT   \\
\enddata
\tablecomments{
(1) Object name;
(2) Spectroscopic z;
(3) Nucleus apparent R magnitude (*: obtained from spectral decomposition);
(4) EW of CaII $\lambda$3934;
(5) Estimated z;
(6) Error on estimated z;
(7) Telescope used for observations.
}
\end{deluxetable}

\end{document}